\documentclass[aps,preprintnumbers,superscriptaddress,longbibliography]{revtex4-1}
\usepackage{amsmath, mathrsfs, amssymb,amsfonts,amsthm,graphicx, epsf, dcolumn, yfonts}
\usepackage[normalem]{ulem}
\usepackage[hyperfootnotes=true]{hyperref}
\usepackage{color}
\usepackage{amssymb}
\usepackage{slashed}
\usepackage{setspace}
\usepackage{cancel}
\usepackage{wasysym}
\usepackage{tikz-feynman}
\usepackage{float}
\usepackage[utf8]{inputenc}
\usepackage{todonotes}
\usepackage{tikz}
\usepackage{tikz-cd}
\usepackage{feynmf}
\usepackage{subcaption}
\usepackage{array}
\usepackage{tabularray}
\usepackage{tabularx}
\usepackage{ragged2e}
        \hypersetup{
           breaklinks=true,   
           colorlinks=true,   
           pdfusetitle=true,  
        }        
\pdfoutput=1
\newcommand{\lin}{L}
\newcommand{\rind}{+}
\newcommand{\lind}{-}
\newcommand{\q}{q}
\newcommand{\p}{p}
\newcommand{\xq}{\phi}

\newcommand{\xql}{\phi^-}
\newcommand{\xqr}{\phi^+}

\def\<{\langle}
\def\>{\rangle}

\newcommand{\cqstate}{\varrho}

\newcommand{\be}{\begin{eqnarray} \begin{aligned}}
\newcommand{\ee}{\end{aligned} \end{eqnarray} }
\newcommand{\benn}{\begin{eqnarray*} \begin{aligned}}
\newcommand{\eenn}{\end{aligned} \end{eqnarray*} }

\newcommand{\ben}{\begin{eqnarray} \begin{aligned}}
\newcommand{\een}{\end{aligned} \end{eqnarray} }

\newcommand{\bc}{\begin{center}}
\newcommand{\ec}{\end{center}}

\newcommand{\Tr}{\mathop{\mathsf{Tr}}\nolimits}

\newcommand{\e}{\mathrm{e}}

\newcommand{\beq}{\begin{eqnarray} \begin{aligned}}
\newcommand{\eeq}{\end{aligned} \end{eqnarray} }
\newcommand{\bea}{\begin{array}}
\newcommand{\eea}{\end{array}}

\newcommand{\bee}{\begin{enumerate}}
\newcommand{\eee}{\end{enumerate}}
\newcommand{\bei}{\begin{itemize}}
\newcommand{\eei}{\end{itemize}}

\usepackage{amsfonts}

\def\01{\{0,1\}}

\newcommand{\cE}{\mathcal{E}}

\newcommand{\new}[1]{{\textcolor{black}{#1}}}
\newcommand{\jono}[1]{\textcolor{black}{#1}}

\newcommand{\str}{{\vec{x}}}

\def\N{\mathcal{N}}
\def\<{\langle}
\def\>{\rangle}

\def\s{\,\,\,\,}

\newtheorem*{rep@theorem}{\rep@title}
\newcommand{\newreptheorem}[2]{%
\newenvironment{rep#1}[1]{%
 \def\rep@title{#2 \ref{##1} (restatement)}%
 \begin{rep@theorem}}%
 {\end{rep@theorem}}}
\makeatother

\newreptheorem{thm}{Theorem}
\newreptheorem{lem}{Lemma}

\newcommand{\zach}[1]{{\color{black}#1}}

\def\T{\bf T}

\def\z{{ z}}

\def\g{{g}}

\def\T00{{\bf T_{NN}}}

\def\0mom{{\bar{\Gamma}^{\alpha\beta}(\z)}}
\def\1mom{{\Gamma^{\alpha\beta}_1(\z)}}
\def\2mom{{\Gamma^{\alpha\beta}_2(\z)}}

\tikzset{
my loop/.style={to path={
.. controls +(50:1.1) and +(130:1.1) .. (\tikztotarget) \tikztonodes}}}
\tikzset{
my loop2/.style={to path={
.. controls +(50:0.8) and +(130:0.8) .. (\tikztotarget) \tikztonodes}}}
\tikzset{
my loop3/.style={to path={
.. controls +(10:0.8) and +(95:0.8) .. (\tikztotarget) \tikztonodes}}}
\tikzset{
my loop4/.style={to path={
.. controls +(-50:0.5) and +(10:0.4) .. (\tikztotarget) \tikztonodes}}}
\tikzset{
my loop5/.style={to path={
.. controls +(230:0.5) and +(170:0.4) .. (\tikztotarget) \tikztonodes}}}

\long\def\diagramone{\begin{tikzpicture}
\draw[fill=black, dashed](0,0)circle(2pt)  node[anchor=north east]{$\phi^+$}   -- +(2cm,0) circle(2pt)node[anchor=north west]{$\phi^+$};
\path (0.9cm,-0.2cm) node[anchor=north]{ $-\frac{i\hbar}{m_{\phi}^2}$} (0cm,+0.1cm) ;
\end{tikzpicture}\hspace{0.5cm}
\begin{tikzpicture}
\draw[fill=black, dotted](0,0)circle(2pt)  node[anchor=north east]{$\phi^-$}   -- +(2cm,0) circle(2pt)node[anchor=north west]{$\phi^-$};
\path (1.1cm,-0.2cm) node[anchor=north]{ $\frac{i\hbar}{m_{\phi}^2}$} (0cm,+0.1cm) ;
\end{tikzpicture}\hspace{0.5cm}
\begin{tikzpicture}
\draw[fill=black](0,0)circle(2pt)  node[anchor=north east]{$q$}   -- +(2cm,0) circle(2pt)node[anchor=north west]{$q$};
\path (1.1cm,-0.2cm) node[anchor=north]{ $\frac{D_2}{m_q^4}$} (0cm,+0.1cm) ;
\end{tikzpicture} 
}
\long\def\interactiondiagram{\begin{tikzpicture}
\coordinate (A) at (1,0);
\node[xshift=0.8cm] at (A) {$-\frac{\lambda m_q^2}{2D_2}$};
\draw[fill=black, dashed](0,0)   -- +(1cm,0) circle(2pt)node[anchor=north west]{} -- (1.5cm,0.865cm);
\draw[fill=black](1cm,0) -- (1.5cm,-0.865cm);
\end{tikzpicture}\hspace{0.5cm}
\begin{tikzpicture}
\coordinate (A) at (1,0);
\node[xshift=0.8cm] at (A) {$-\frac{\lambda m_q^2}{2D_2}$};
\draw[fill=black, dotted](0,0)   -- +(1cm,0) circle(2pt)node[anchor=north west]{} -- (1.5cm,0.865cm);
\draw[fill=black](1cm,0) -- (1.5cm,-0.865cm);
\end{tikzpicture}}

\long\def\interactiondiagramsix{\begin{tikzpicture}
\coordinate (A) at (2,0);
\node[xshift=0.8cm] at (A) {$-\frac{12 \lambda^2 }{D_2}$};
\draw[fill=black, dotted](0cm,0)   -- + (1cm,0) circle(2pt)node[anchor=north west]{};
\draw[fill=black](1cm,0) -- (0.29cm,0.71cm);
\draw[fill=black,dotted](1cm,0) -- (1.71cm,-0.71cm);
\draw[fill=black](1cm,0) -- (1.71cm,0.71cm);
\draw[fill=black,dotted](1cm,0) -- (0.29cm,-0.71cm);
\draw[fill=black, dotted](1cm,0) -- (2cm,0cm);
\end{tikzpicture}\hspace{0.5cm}
\begin{tikzpicture}
\coordinate (A) at (2,0);
\node[xshift=0.8cm] at (A) {$-\frac{12 \lambda^2 }{D_2}$};
\draw[fill=black, dotted](0cm,0)   -- + (1cm,0) circle(2pt)node[anchor=north west]{};
\draw[fill=black](1cm,0) -- (0.29cm,0.71cm);
\draw[fill=black,dotted](1cm,0) -- (1.71cm,-0.71cm);
\draw[fill=black,dotted](1cm,0) -- (1.71cm,0.71cm);
\draw[fill=black,dotted](1cm,0) -- (0.29cm,-0.71cm);
\draw[fill=black](1cm,0) -- (2cm,0cm);
\end{tikzpicture}\hspace{0.5cm}
\begin{tikzpicture}
\coordinate (A) at (2,0);
\node[xshift=0.8cm] at (A) {$-\frac{12 \lambda^2 }{D_2}$};
\draw[fill=black, dotted](0cm,0)   -- + (1cm,0) circle(2pt)node[anchor=north west]{};
\draw[fill=black](1cm,0) -- (0.29cm,0.71cm);
\draw[fill=black](1cm,0) -- (1.71cm,-0.71cm);
\draw[fill=black,dotted](1cm,0) -- (1.71cm,0.71cm);
\draw[fill=black,dotted](1cm,0) -- (0.29cm,-0.71cm);
\draw[fill=black,dotted](1cm,0) -- (2cm,0cm);
\end{tikzpicture}
}
\long\def\interactiondiagramsixdash{\begin{tikzpicture}
\coordinate (A) at (2,0);
\node[xshift=0.8cm] at (A) {$-\frac{12 \lambda^2 }{D_2}$};
\draw[fill=black, dashed](0cm,0)   -- + (1cm,0) circle(2pt)node[anchor=north west]{};
\draw[fill=black](1cm,0) -- (0.29cm,0.71cm);
\draw[fill=black,dashed](1cm,0) -- (1.71cm,-0.71cm);
\draw[fill=black](1cm,0) -- (1.71cm,0.71cm);
\draw[fill=black,dashed](1cm,0) -- (0.29cm,-0.71cm);
\draw[fill=black, dashed](1cm,0) -- (2cm,0cm);
\end{tikzpicture}\hspace{0.5cm}
\begin{tikzpicture}
\coordinate (A) at (2,0);
\node[xshift=0.8cm] at (A) {$-\frac{12 \lambda^2 }{D_2}$};
\draw[fill=black, dashed](0cm,0)   -- + (1cm,0) circle(2pt)node[anchor=north west]{};
\draw[fill=black](1cm,0) -- (0.29cm,0.71cm);
\draw[fill=black,dashed](1cm,0) -- (1.71cm,-0.71cm);
\draw[fill=black,dashed](1cm,0) -- (1.71cm,0.71cm);
\draw[fill=black,dashed](1cm,0) -- (0.29cm,-0.71cm);
\draw[fill=black](1cm,0) -- (2cm,0cm);
\end{tikzpicture}\hspace{0.5cm}
\begin{tikzpicture}
\coordinate (A) at (2,0);
\node[xshift=0.8cm] at (A) {$-\frac{12 \lambda^2 }{D_2}$};
\draw[fill=black, dashed](0cm,0)   -- + (1cm,0) circle(2pt)node[anchor=north west]{};
\draw[fill=black](1cm,0) -- (0.29cm,0.71cm);
\draw[fill=black](1cm,0) -- (1.71cm,-0.71cm);
\draw[fill=black,dashed](1cm,0) -- (1.71cm,0.71cm);
\draw[fill=black,dashed](1cm,0) -- (0.29cm,-0.71cm);
\draw[fill=black,dashed](1cm,0) -- (2cm,0cm);
\end{tikzpicture}
}

\begin{document}

\title{Covariant path integrals for quantum fields back-reacting on classical space-time}

\author{Jonathan Oppenheim}

\affiliation{Department of Physics and Astronomy, University College London, Gower Street, London WC1E 6BT, United Kingdom}
\author{Zachary Weller-Davies}

\affiliation{Perimeter Institute for Theoretical Physics, Waterloo, Ontario, Canada}

\begin{abstract}
\noindent We introduce configuration space path integrals for quantum fields interacting with classical fields. We show that this can be done consistently by proving that the dynamics are completely positive directly, without resorting to master equation methods. These path integrals allow one to readily impose space-time symmetries, including Lorentz invariance or diffeomorphism invariance.  They generalize and combine the Feynman-Vernon path integral of open quantum systems and the stochastic path integral of classical stochastic dynamics while respecting symmetry principles.  We introduce a path integral formulation of general relativity where the space-time metric is treated classically. The theory  is a candidate for a fundamental theory that reconciles general relativity with quantum mechanics. \jono{The theory is manifestly covariant, and may be inequivalent to the theory derived using master-equation methods.} \new{We prove that entanglement cannot be created via the classical field, reinforcing proposals to test the quantum nature of gravity via entanglement generation.}
\end{abstract}

\maketitle

\noindent \textbf{Introduction.}
Effective theories are ubiquitous in physics: from particle physics to classical statistical mechanics, we often make approximations to an underlying physical theory in order to simplify the dynamical description at hand. Broadly, there are two approaches to constructing effective field theories \cite{GeorgiEFT}. In the Wilsonian approach \cite{Wilson, GeorgiEFT}, one starts with a high energy theory and asks how the effective low energy description changes as high momentum modes are integrated out. The second approach involves modifying the theory by hand in order to isolate the desired degrees of freedom whilst still trying to keep phenomenological accuracy. This usually involves modifying the high-energy description so that the effective description is simpler and easier to use.

Often, we are interested in the effective description of a system where one part behaves classically and the other quantum mechanically, so the system should be described by an effective theory of combined classical-quantum (CQ) dynamics. We do this when we model a quantum measurement since we treat the measurement device as being effectively classical. Of particular interest is where the gravitational field is treated as classical while matter remains quantum since we currently have no quantum theory of gravity. Here, we are often interested in the regime where the quantum fields back-react onto the geometry, for example when studying vacuum fluctuations in inflationary cosmology or black hole evaporation. Since gravity is a field theory, the correct description of quantum matter back-reacting on classical space-time should be described by an effective field theory where the matter degrees of freedom are quantum mechanical, and the gravitational field is treated as being effectively classical.

Consistent classical-quantum (CQ) master equations, such as the examples introduced in \cite{blanchard1995event,diosi1995quantum},  have been used to study the interaction between classical and quantum systems from a master equation perspective. Consistent CQ master equations have been studied in a variety of different contexts \cite{alicki2003completely, poulinKITP2,Disi2017OnGD, Oppenheim:2018igd,oppenheim2020objective}, including gravity \cite{Diosi:2011vu, Kafri_2014,tilloy2016sourcing, Oppenheim:2018igd,pqconstraints,dec_Vs_diff}, and can be shown to be completely positive, trace preserving (CPTP), and preserve the split between classical and quantum degrees of freedom. CPTP dynamics is required in order to preserve the statistical properties of the density matrix since probabilities are positive and sum to one. The most general form of CPTP classical-quantum dynamics has also been derived \cite{Oppenheim:2018igd,UCLPawula}, and takes an analogous form to that of the GSKL or Lindblad equation \cite{GSKL,Lindblad:1975ef} in open quantum systems and the rate equation in stochastic classical dynamics \cite{gardiner2004handbook}.

  Importantly, the resulting dynamics do not suffer from the same problems as the standard semi-classical equations \cite{moller1962theories, rosenfeld1963quantization}. There, the back-reaction on the classical degrees of freedom is sourced by an expectation value of the quantum state and is known to be inconsistent, inducing a break-down of either operational no-signaling, the Born rule, or composition of quantum systems under the tensor product \cite{Gisin:1989sx,eppley1977necessity,tilloy2016sourcing,galley2021nogo, page1981indirect}. Instead, the CQ dynamics we consider here is the most general form of dynamics consistent with the state space of quantum mechanics and is both quantitatively and qualitatively different from the standard semi-classical equations \cite{UCLHealing}.

  However, it is not known how to frame classical-quantum dynamics in a manifestly covariant framework, which places serious restrictions on any dynamics with an effectively classical field. The problem with the master equation picture is twofold. Firstly, from a practical point of view, field theories are generally better suited to path integral methods. Secondly, in a master equation picture, it is difficult to impose symmetries directly on the master equation. Indeed, writing down master equations for classical-quantum fields directly, without knowing whether or not they are covariant or uphold space-time symmetries, seems to go against much of the principles of modern physics, where one starts with actions based on symmetry principles. If one took the position that there is a fundamentally classical field, such as the gravitational field, it is also not obvious how one could couple it to the standard model whilst simultaneously ensuring symmetry principles and renormalizability are upheld.  

In an accompanying paper \cite{UCLPILONG}, we give an in-depth analysis of classical-quantum master equations and their associated path integrals, the results of which motivated much of the present work. This work introduces a fully covariant path integral framework to study classical fields interacting with quantum ones. We prove the dynamics are completely positive directly from the path integral perspective, and we do not resort to master equation methods. This is especially important since, in general, it is only possible to go from a master equation picture to a path integral picture when the master equation is less than quadratic in classical or quantum momenta \cite{UCLPILONG}. We find a family of CQ path integrals that are generated by an action, so it is easy to write down theories with space-time symmetries, including gauge symmetries. We do not study the renormalization properties of the dynamics explicitly, \jono{however, since this work appeared on the arxiv, we have since found that the pure gravity theory presented here is formally renormalizable\cite{grudka2023renorm}.  }

The path integral we study is a generalization of the Feynman path integral for quantum systems, and the stochastic path integral used to study classical stochastic processes \cite{Onsager1953Fluctuations,freidlin1998random} (see Table \ref{tab: pathintegrals} for a comparison). It combines these forms and includes an interacting term between the classical and quantum fields. When there is no back-reaction on the classical field, the path integral reduces to standard quantum theory with an action that depends on a classical variable. 
When there is back-reaction on the classical field, the path integral includes a summation over all classical configurations and gives rise to a natural, and in fact, necessary \cite{dec_Vs_diff} mechanism for decoherence. Though the quantum state decoheres, the path integrals preserve purity on the quantum system so that quantum states are mapped to quantum states. There is no loss of quantum information: this is a feature of CQ dynamics under certain natural conditions, which can lead to pure quantum state trajectories conditioned on the trajectory of the classical degree of freedom \cite{oppenheim2020objective, UCLHealing}. As discussed in \cite{blanchard1995event, Oppenheim:2018igd,oppenheim2020objective, UCLHealing} such dynamics do not require the Born rule to explain the state update rule and probabilistic outcomes of measurements if the classical field is taken as fundamental. Unlike spontaneous collapse models \cite{Pearle:1988uh,PhysRevA.42.78,BassiCollapse,PhysRevD.34.470, GisinCollapse, relcollapsePearle, 2006RelCollTum}, the classical variable is itself taken to be dynamical.

Our results have consequences for any theory with a degree of freedom that behaves classically, whether effective or fundamental. With this in mind, we provide a possible template for studying CQ field theories, and we introduce a class of classical-quantum actions which can be used to construct theories with a sensible classical limit. The corresponding path integral can be understood in terms of summing over all classical and quantum paths, where the classical paths deviating too much from their semi-classical configuration are suppressed by a coupling $D_{0}$, which also governs the strength of the quantum decoherence. \jono{We give an explicit Lorentz invariant model of a classical field coupled to a quantum field. We show how to use perturbation theory to compute correlation functions in Appendix \ref{sec: peturbative}. This can then be used to compute vacuum expectation values which place experimental constraints on the theory. We also include a discussion of normalization techniques in Appendix \ref{app: normalizing}.}  Since we do not have a full theory of quantum gravity, of particular importance is the construction of theory of quantum matter back-reacting on classical space-time, and we discuss the application of our work to the gravitational setting, giving an example of a CQ theory of gravity which gives the trace of Einstein's equations on average. We also construct a path integral for the full set of Einstein's equations. We have since shown \new{that this path integral gives the correct classical limiting behavior. In particular  we have since shown in \cite{grudka2023renorm} that fluctuations away from the Newtonian potential and scalar mode are suppressed by showing that the scalar two-point function is positive semidefinite. The two-point function of the tensor mode, which capture the dynamical degrees of freedom of general relativity has also been show to be positive semidefinite\cite{oppenheim2025tensormode}}. \new{We show that the classical gravitational field cannot generate entanglement, demonstrating that this provides a witness for a quantum theory of gravity\cite{bose2017spin,marletto2017gravitationally}. This resolves a question which has seen considerable debate\cite{danielson2022gravitationally,husain2022dynamics,ma2022limits,fragkos2022inference,christodoulou2023locally,gollapudi2025state,martin2023gravity,trillo2025diosi}}. The theory can be considered as a fundamental theory which is an alternative to quantum gravity. There is also a regime where it may be an effective description of a fully quantum theory of gravity, after taking the classical-quantum limit as outlined in \cite{layton2024classical}.

\noindent \textbf{Classical-quantum dynamics.} We first introduce the general formalism used to describe a classical degree of freedom coupled to a quantum one, and we denote a generic classical degree of freedom by $\z$. For example, it could be a classically treated position variable $\z=q$, or a point in phase space $\z=(q,p)$. When one considers a hybrid system, the natural set of states to consider are hybrid classical-quantum (CQ) states. Formally, a classical-quantum state associates to each classical variable an un-normalized density matrix $\cqstate(\z,t) =p(\z,t) \sigma(\z,t)$ such that $\Tr_{\mathcal{H}}{\cqstate(\z)} = p(\z,t) \geq 0$ is a normalized probability distribution over the classical degrees of freedom and $\int d\z \cqstate(\z,t) $ is a normalized density operator on a Hilbert space $\mathcal{H}$. Intuitively, $p(\z,t)$ can be understood as the probability density of being in the phase space point $\z$ and  $\sigma(\z,t)$ as the \textit{normalized} quantum state one would have given the classical state $\z$ occurs.

Classical-quantum dynamics can then be understood as the set of linear dynamics which maps CQ states to CQ states. Linearity is required in order to maintain a probabilistic interpretation of the density matrix. The dynamics must be completely positive since we require that states be mapped to other states even when the dynamics act on half an entangled quantum state. In analogy with Krauss theorem for quantum operations, the most general form of CP dynamics mapping CQ states onto themselves is described by \cite{Oppenheim:2018igd, UCLPawula} \begin{equation}\label{eq: CPmap}
 \cqstate(\z,t+ \delta t) = \int d\z' \Lambda^{\mu\nu}(\z|\z',\delta t) \lin_{\mu}\cqstate(\z',t) \lin_{\nu}^{\dag},
\end{equation}  
where the $ \Lambda^{\mu \nu}(\z|\z',\delta t)$ defines a positive matrix-measure in $z,z'$. In Equation \eqref{eq: CPmap}, the operators $\lin_{\mu}$ are an arbitrary set of operators on the Hilbert space, and normalization of probabilities requires
\begin{equation}\label{eq: prob}
\int d\z  \Lambda^{\mu\nu}(\z|\z',\delta t) \lin_{\nu}^{\dag}\lin_{\mu} =\mathbb{I}.
\end{equation}

When the dynamics are Markovian, completely positive CQ master equations can be derived from Equation \eqref{eq: CPmap} and have been studied in \cite{alicki2003completely, poulinKITP2, Oppenheim:2018igd,oppenheim2020objective, Diosi:2011vu, Oppenheim:2018igd,pqconstraints,dec_Vs_diff}. However, it is well known that it is only possible to go from a master equation approach to a position space path integral approach when the master equation is at most quadratic in momenta, or else one cannot perform the Gaussian path integral exactly. Therefore any method of proving consistent dynamics without master equation methods is useful. We shall here work \textit{entirely} within a path integral framework without resorting to master equation methods, and we shall prove complete positivity of the dynamics directly from the path integral approach. 

The path integral tells us how the components of the density matrix evolve. Including a classical variable $q$, the path integral should tell us how to evolve the components of a classical-quantum state 
\begin{equation}\label{eq: statecomponents}
    \cqstate( q,t) = \int d \phi^+ d\phi^- \cqstate( q,t,\phi^+,\phi^-) \ | \phi^+ \rangle \langle \phi^- |,
\end{equation}
where $\phi$ represents a continuous quantum degree of freedom and  $\cqstate( q,t,\phi^+,\phi^-) = \langle \phi^+| \rho(q,t)|\phi^- \rangle $ are the components of the CQ state. Writing \eqref{eq: statecomponents} out explicitly, generically, a path integral will take the form
\begin{equation}\label{eq: transition0}
    \rho(q,\phi^+,\phi^-t_f)  = 
    \int \mathcal{D}q \mathcal{D} \phi^+ \mathcal{D} \phi^- \N e^{\mathcal{I}[q,\phi^+,\phi^-,t_i,t_f]}   \rho(q_i,\phi^+_i,\phi^-_i, t_i).
\end{equation}
In Equation \eqref{eq: transition0} it is implicitly understood that boundary conditions are to be imposed at $t_f$, and we have included a normalization factor $\N$ which can depend on the point in configuration space at each instance \cite{UCLNormPI}. In the purely quantum case, one has $\mathcal{I}[\phi^+,\phi^-,t_i,t_f] = i S[\phi^+, t_i, t_f] - i S[\phi^-, t_i, t_f]$ and the path integral is doubled since we are considering density matrices so we must sum over all bra and ket paths.

When the action contains higher derivatives, we can also include additional initial conditions on the time derivatives of the fields in \eqref{eq: transition0} \cite{Hawking:2001yt}. 

\noindent \textbf{Main result: \jono{A Completely Positive Norm preserving Path Integral.}} Having introduced the classical-quantum formalism, let us now state and prove our main result:
{\it Any time-local classical-quantum path integral with action of the form 
 \begin{equation}\label{eq: positiveCQ}
  \mathcal{I}( \xql,  \xqr,  \q,t_i,t_f) =  \mathcal{I}_{CQ}(\q,\phi^+, t_i,t_f)\\
   + \mathcal{I}^*_{CQ}(\q,\phi^-, t_i,t_f)  - \mathcal{I}_C(\q, t_i,t_f)
  + \int_{t_i}^{t_f} dt \sum_{ \gamma} c^{\gamma}(q,t)( L_{\gamma}[\phi^+] L^{*}_{\gamma}[\phi^-])
\end{equation}
 defines completely positive CQ dynamics when the terms in Equation \eqref{eq: positiveCQ} have the following properties: $L_{\gamma}[\phi^{\pm}]$ can be any functional of the bra and ket variables, $c^{\gamma}\geq 0$, $\mathcal{I}_{C}$ is positive (semi) definite, and the real part of $\mathcal{I}_{CQ}$ is negative (semi) definite. We implicitly assume that $c^{\gamma}$ is chosen so that the path integral converges. }  

 In the field-theoretic case, the final line of Equation \eqref{eq: positiveCQ} is replaced by 
 \begin{equation}
     \int_{t_i}^{t_f} dx \sum_{ \gamma} c^{\gamma}(q,x)( L_{\gamma}[\phi^+] (x) L^{*}_{\gamma}[\phi^-](x)),
 \end{equation} and the resulting path integral in Equation \eqref{eq: positiveCQ} will be CP so long as $c^{\gamma}(q,x)$ is positive. 
 
 In Equation \eqref{eq: positiveCQ} $\mathcal{I}_{CQ}$ determines the CQ interaction on each of the ket and bra paths and $\mathcal{I}_C(\q, t_i,t_f)$ is a purely classical action which takes real values. The above requirements on positive defniteness have been imposed in order for the path integral to be convergent. This condition also arises when studying path integrals associated to CQ master equations \cite{UCLPILONG}. For example, one can take the classical action $\mathcal{I}_C$ to be the action associated to the path integral of the Fokker-Planck equation \eqref{eq: FPaction} \cite{Onsager1953Fluctuations,Kleinert, Weber_2017} which must be positive (semi) definite in order for the path integral to converge. The term on the final line of Equation \eqref{eq: positiveCQ} contains cross terms between the bra and ket branches $\phi^+$, $\phi^-$ which sends pure states to mixed states and corresponds to including additional noise in the dynamics. It takes the form of a Krauss map acting on the CQ state, which is what ensures complete positivity, and allows one to include classical-quantum Feynman-Vernon \cite{FeynmanVernon, Baidya:2017eho} terms into the action.
  
  If all the $c^{\gamma}=0$, the $\phi^+$ and $\phi^-$ integrals factorize in Equation \eqref{eq: positiveCQ}, the path integral preserves the purity of the quantum state conditioned on the classical trajectory. \jono{This can be seen from the fact that the absence of $\phi^+\phi^-$ couplings, mean that the bra field evolves independently of the ket field. If initially at $q$, the quantum system is in state $|\phi\rangle\langle\phi|$ the first term in Eq. \eqref{eq: positiveCQ} will evolve $|\phi\rangle$ to another pure state $|\phi(t)\rangle$, while the second term in Eq. \eqref{eq: positiveCQ} will evolve $\langle\phi|$ to the pure state $\langle\phi|$, leaving the final density matrix in pure state $|\phi(t)\rangle\langle\phi(t)|$.} In this case, the absence of cross terms in the action, despite the requirement of Lindblad terms in the hybrid master equation \cite{dec_Vs_diff}, is a remarkable consequence of saturating the decoherence vs diffusion trade-off \cite{UCLPILONG}. 
 
\jono{To see this more clearly,} it is useful to split $\mathcal{I}_{CQ}$ into its real and imaginary components $\mathcal{I}_{CQ} =\mathcal{R}_{CQ} + i \mathcal{S}_{Q}$. Then Equation \eqref{eq: positiveCQ} (with $c_n=0$) reads 
 \begin{equation}\label{eq: decomp}
  \mathcal{I}^{\pm} = \mathcal{R}_{CQ}^+ + \mathcal{R}_{CQ}^- + i (\mathcal{S}_{CQ}^+ - \mathcal{S}_{CQ}^-) - \mathcal{I}_C,
 \end{equation}
and we are able to get some intuition for each term. Heuristically expanding the actions, or more properly their Lagrangian's, in terms of their field dependence $\mathcal{S}_{CQ} \sim \sum_{m} a_m(q)s_m(\phi)$ and $\mathcal{R}_{CQ} \sim \sum_{m} b_m(q)r_m(\phi)$ we see that 
 \begin{equation}
     \begin{split}
        &  \mathcal{S}_{CQ}^+ - \mathcal{S}_{CQ}^- \sim \sum_m a_m(q)(s_m(\phi^+) - s_m( \phi^-)) \\
        & \mathcal{R}_{CQ}^+ + \mathcal{R}_{CQ}^- \sim \sum_m b_m(q) (r_m(\phi^+) + r_m(\phi^-)).
     \end{split}
 \end{equation}
 Hence, the imaginary part of the integral is associated with things like coherence, which depend on the difference between the ket and bra components of the density matrix, whilst the real part of the action depends on the sum of the left and right components on the density matrix, which are things like its expectation value. Moreover, conditioned on a classical trajectory $\bar{q}(t)$ - which can be represented by inserting a delta function $\delta(q(t)-\bar{q}(t))$ into the classical part of the path integral - we see that the evolution of the quantum state factorizes between the $\phi^{\pm}$ integrals and hence keeps pure quantum states pure. We shall largely focus on this case; it can be shown that any CQ dynamics which does not preserve the purity of the quantum state conditioned on the classical degree of freedom can be embedded into a larger classical space where the quantum state remains pure, in a CQ version of purification \cite{UCLHealing}.
 
The back-reaction of the quantum system on the classical one is contained in the real components of the CQ action $\mathcal{R}^{\pm}_{CQ}$. Indeed, when $\mathcal{R}^{\pm}_{CQ}=0$, the path integral in Equation \eqref{eq: decomp} reduces to the standard quantum path integral for the density matrix but also includes a classical variable which can undergo its own autonomous dynamics due to the inclusion of the classical action $\mathcal{I}_C$. However, whenever there is back-reaction, Equation \eqref{eq: positiveCQ} \textit{necessarily} describes non-unitary evolution: this was proved generally in \cite{dec_Vs_diff} using master equation methods.
 
To prove that the dynamics described by Equation \eqref{eq: positiveCQ} gives rise to consistent CQ dynamics, we must first show that it leads to completely positive dynamics preserving the positivity of the CQ state. 
Recall that positivity of the CQ state means that for any Hilbert space vector $|v(q)\rangle $ we have $\Tr[|v(q)\rangle \langle v(q)| \cqstate(q)] \geq 0 $. In components, complete positivity is equivalent to asking that for any vector $|v(q)\rangle$ with components $v(\phi, q) = \langle \phi|v(q) \rangle$ we have
\begin{equation}\label{eq: CPpath}
    \int d\phi^{+} d\phi^- v(\phi^+,q)^* \cqstate(\phi^+,\phi^-,\q) v(\phi^-,q) \geq 0 .
\end{equation}
A CQ dynamics $\Lambda$ is said to be positive if it preserves the positivity of CQ states and completely positive if $\mathbb{I} \otimes \Lambda$ is positive when we act with the identity on any larger system.

 Since we assume the dynamics are time-local, we can perform a short-time expansion of the path integral. 
For the action in Equation \eqref{eq: positiveCQ}, in Appendix \ref{sec: proofofpos} we show that the path integral integrand always factorizes into the form 
\begin{equation}\label{eq: positivelooking} [ e^{\mathcal{I}^+[\phi^+,q]}e^{\mathcal{I}^-[\phi^-,q]^*} +  \delta t \sum_{\gamma} c^{\gamma}(L^+_{\gamma} e^{\mathcal{I}^+[\phi^+,\q]}) ( L^-_{\gamma} e^{\mathcal{I}^-[\phi^-,\q]})^* ] \e^{-\mathcal{I}_C[q]} + \dots,\end{equation}
 Because Equation \eqref{eq: positivelooking} factorizes between $\pm$ branches, it is manifestly completely positive, which can be seen from the definition of complete positivity in Equation \eqref{eq: CPpath}. It is important to note that because of the exponentials, Equation \eqref{eq: positivelooking} is always strictly positive, meaning that we do not encounter zero norm states. Instead, the problem of negative norm states and ghosts is mapped to the problem of convergence of the path integral \cite{Hawking:2001yt}. 

\zach{The path integral defined by Equation \eqref{eq: positiveCQ} is completely positive, and the other requiremnt is that it be  norm  preserving. For time local dynamics, it is always possible to normalize a CP map in a linear manner to arrive at a CP norm preserving dynamics. Specifically, any time-local CP CQ map of the form
\begin{equation}
    \frac{\partial \cqstate(z)}{\partial t} = \int d z^{\prime} W^{\mu \nu}\left(z | z^{\prime}\right) L_{\mu} \cqstate\left(z^{\prime}\right) L_{\nu}^{\dagger} 
\end{equation} can be normalized by subtracting $\frac{1}{2}\int d z^{\prime} W^{\mu \nu}\left(z^{\prime} | z\right) \{L_{\nu}^{\dagger} L_{\mu}, \cqstate(z)\}_+$ to yield CP norm preserving dynamics \cite{Oppenheim:2018igd}.

With this in mind, for time-local dynamics, Equation \eqref{eq: positiveCQ} can always be normalized and taking this into account we can include a normalization factor in the CQ path integral. In practice, and for the path integrals we introduce in this work, normalization of the path integral is accounted for through the classical and quantum kinetic terms in the action \cite{Weller-Davies:2024zcb} (see Appendix \ref{app: normalizing}).}

\noindent \textbf{Comparison to classical path integrals.} The path integral action we introduce in Equation \eqref{eq: positiveCQ} is general. Therefore, it is useful to find CQ actions that give rise to dynamics that have a sensible physical interpretation. To gain some intuition for the classical part of path integral, we can consider the Fokker-Plank Equation for a classical probability density $p(z)$ 
\begin{equation}\label{eq: FP}
    \frac{\partial p(z)}{\partial t} = -\frac{\partial}{\partial z_i}[D_{1,i}(z) p(z)] + \frac{1}{2}\frac{\partial^2}{\partial z_i \partial z_j}[D_{2,ij}(z) p(z) ],
\end{equation}
where $z=(q_1,p_1,\dots, q_n, p_n)$ for an $n$ dimensional system \cite{risken1989fpe}.

In \eqref{eq: FP} the coefficient $D_{1,i}$ characterizes the amount of \textit{drift} in the system, and is equal to the evolution of the expectation value of $z$, $\partial_t{\langle z_i \rangle }$. If $D_{1,q_i}$ also depends on $p_i$ then it contributes a  friction term. The matrix $D_{2,ij}$ characterizes the amount of \textit{diffusion} in the system and is equal to $\partial_t \langle z_i z_j \rangle $.
The corresponding path integral is given by \cite{Onsager1953Fluctuations,freidlin1998random,Weber_2017, Kleinert}
\begin{equation}
    p(z,t_f) = \int \mathcal{D}z \N e^{-\mathcal{I}_C(z,t_i,t_f)}p(z_i,t_i),
\end{equation}
where 
\begin{equation}\label{eq: FPaction}
    \mathcal{I}_C(z,t_i,t_f) = \frac{1}{2} \int_{t_i}^{t_f} dt [\frac{dz_i}{dt}-D_{1,i}(z)]D_{2,ij}^{-1}[\frac{dz_j}{dt}-D_{1,j}(z)].
\end{equation}
When the matrix $D_2$ doesn't depend on $z$, the normalization is easily computated since the integral is a Gaussian. In this case it doesn't depend on configuration space and can be taken outside the path integral.
The path integral has a natural interpretation in terms of suppressing classical paths which deviate from their expected drift $D_1$ by an amount that depends on the inverse of the diffusion coefficient $D_2^{-1}$. If $D_2$ is $z$ dependent, Equation \eqref{eq: FPaction} can also contain an anomalous contribution \cite{UCLPILONG}, but we shall not include it here since \eqref{eq: FPaction} still defines positive classical dynamics.

The simplest non-trivial case is where one diffuses only in momenta. In this case, $\dot{q}_i = \frac{p_i}{m_i}$ and the momentum integral acts to enforce a delta function over $\delta(p_i -m_i \dot{q}_i)$. Integrating out the momentum variables, the result is a path integral over only the configuration space variables $q_i$ with action 
\begin{equation}\label{eq: actionClass}
  \mathcal{I}_C(q,t_i,t_f) = 
  \frac{1}{2} \int_{t_i}^{t_f} dt [m_i\frac{d^2q_i}{dt^2}-D_{1,i}(q)]D_{2,ij}^{-1}[m_j\frac{d^2q_j}{dt^2}-D_{1,j}(q)],
\end{equation}
from which we see that the path integral acts to suppress paths away from their expected equation's of motion with the amount depending on $D_2$. 

Taking the expected classical equation of motion to itself be generated by an action $S_C$, the action in \eqref{eq: actionClass} can be re-written as 
\begin{equation}\label{eq: lagClass}
  \mathcal{I}(q,t_i,t_f) = 
  \frac{1}{2} \int_{t_i}^{t_f} dt \frac{\delta S_C}{\delta q_i}D_{2,ij}^{-1}\frac{\delta S_C}{\delta q_j}.
\end{equation}
Since $S_C$ itself appears in the path integral action $\mathcal{I}(q,t_i,t_f)$, we shall henceforth refer to $S_C$ as the classical \textit{proto-action}. It is important to note that, in general, one can, and generally should, include non-Lagrangian friction terms in the path integral, represented by a more general drift coefficient, as in Equation \eqref{eq: actionClass}.

\noindent \textbf{A natural class of CQ dynamics.} The purely classical action in \eqref{eq: lagClass} generalizes to the combined classical-quantum case. A natural class of theories we find are those derivable from a classical-quantum proto-action $W_{CQ}[q,\phi]$: 
\begin{equation}\label{eq: fieldConfigurationSpace2}
      \mathcal{I}( \xql,  \xqr,  \q,t_i,t_f) =    \int_{t_i}^{t_f} dx \bigg[   i\mathcal{L}_Q^+(\q,x)  - i\mathcal{L}_Q^-(\q,x)       
        -\frac{1}{2}\frac{ \delta  \Delta W_{CQ}}{ \delta q_{i}(x)} D_{0,ij}(\q,x) \frac{\delta \Delta W_{CQ}}{ \delta q_{j}(x)} 
         - \frac{1}{2}\frac{\delta \bar{W}_{CQ} }{ \delta q_i(x)} D_{2, ij}^{-1}(\q,x) \frac{\delta \bar{W}_{CQ} }{ \delta q_j(x)}
    \bigg],
\end{equation}
where we take $D_0(q,x), D_2(q,x)$ to be symmetric, positive semi-definite real matricies  \cite{dec_Vs_diff} and we impose the matrix restriction $ 4D_0 \succeq D_2^{-1}$ to ensure the action takes the form of Equation \eqref{eq: positiveCQ}, and hence is completely positive. We show this explicitly in Appendix \ref{sec: protoactioncp}. \zach{One can further show that Equation \eqref{eq: positiveCQ} is normalized so long as the CQ proto action contains classical kinetic terms and $\mathcal{L}_Q$ contains quantum kinetic terms -- precise normalization conditions are summarized through Equation's \eqref{eq: normalizedCQmainLowerAppendix} and \eqref{eq: normalizedCQApp} in Appendix \ref{app: normalizing}.} In Equation \eqref{eq: fieldConfigurationSpace2} $W_{CQ}[\q,\phi]$ is a real classical-quantum proto-action which generates the dynamics, and we have introduced the notation $\bar{W}_{CQ} = \frac{1}{2 }( W_{CQ}[q,\phi^+] + W_{CQ}[q,\phi^-])$ for the $\pm$ averaged proto-action and $\Delta W_{CQ} =  W_{CQ}[q,\phi^-] - W_{CQ}[q,\phi^+]$ for the difference in the proto-action along the $\pm$ branches. 
$\mathcal{L}_Q^{\pm}(\q,x)$ denotes the purely quantum evolution, which can be any quantum field theory Lagrangian density, but should include the coupling terms from $W_{CQ}$ that depend both on $q$ and $\phi^{\pm}$. \jono{In this way, $\mathcal{L}_Q^{\pm}(\q,x)$ generates the evolution of the quantum fields including the action of the classical fields on the quantum fields, while $W_{CQ}$ generates the back-reaction of the quantum fields on the classical fields.}
We will typically take $W_{CQ}[q,\phi^\pm]=\mathcal{L}_Q^\pm(\q,x)$ so that the back-reaction of quantum fields on $\q$ has the same generator as the action of $q$ on the quantum fields, as is the case with two quantum or two classical systems.
 Just as in the classical case, one can also add friction terms to Equation \eqref{eq: fieldConfigurationSpace2} though we shall not do this in the present work. 
 
 For simplicity, we here deal with theories with ''ultra-local'' correlation kernels, 
 meaning the noise kernel is proportional to $\delta^{(3)(}x-y)$
 but we also expect our results to extend to the case where $D_0, D_2$  are positive semi-definite matrix kernels $D_0(x,y), D_2(x,y)$ which have some range \cite{dec_Vs_diff}.

When $4D_0 = D_2^{-1}$, the path integral preserves purity on the quantum system, as shown in \cite{UCLHealing} using master-equation methods. This form of action is motivated by the study of path integrals \cite{UCLPILONG} for CQ master Equations whose back-reaction is generated by a Hamiltonian \cite{Diosi:2011vu, Oppenheim:2018igd, UCLHealing}, as well as the purely classical path integral in Equation \eqref{eq: lagClass}. 
Written in the form of Equation \eqref{eq: fieldConfigurationSpace2}, we see that the action of $D_2$ is to suppress paths that deviate from the $\pm$ averaged Euler-Lagrange equations, which themselves follow from varying the bra-ket averaged proto-action $\bar{W}_{CQ}$, whilst the effect of the $D_0$ term is to decohere the quantum system. The decoherence diffusion trade-off $ 4D_0 \succeq D_2^{-1}$ \cite{dec_Vs_diff}, required for the dynamics to be CP, means that if coherence is maintained for a long time, then there is necessarily lots of diffusion in the classical system away from its most likely path, with the amount depending on both $D_0$ and the strength of the coupling which enters in $W_{CQ}$.

\noindent \textbf{Lorentz invariant classical-quantum dynamics.} Lorentz invariant or covariant pure Linbladians have been studied in \cite{alicki-reldecoherence,poulinKITP2,Baidya:2017eho}.
As a simple example of a classical-quantum Lorentz invariant model, we can consider a classical field $q(x)$ coupled to a quantum field $\phi(x)$ with a manifestly Lorentz invariant proto-action
\begin{equation}\label{eq: Linvariantaction}
    W_{CQ} = \int d^4 x\big[ -\frac{1}{2}\partial_{\mu} q \partial^{\mu}q - \frac{1}{2}m_{q}^2 q^2 
    - \frac{\lambda}{2} q^2 \phi^2\big].
\end{equation}
In this case, assuming $4D_0=D_2^{-1}$, we find the expressions for the CQ coupling terms
\begin{align}\label{eq: decoLor}
   &  \frac{\delta \Delta W_{CQ}}{\delta q}D_0  \frac{\delta \Delta W_{CQ}}{\delta q}  = \lambda^2 D_0 q^2((\phi^+)^2-(\phi^-)^2)^2\\ 
   \label{eq: diffLor}
    &  \frac{\delta \bar{W}_{CQ}}{\delta q}D_2^{-1}  \frac{\delta \bar{W}_{CQ}}{\delta q}  = 4 D_0(\partial^{\mu} \partial_{\mu} q + m_{q}^2 q 
 + \lambda \jono{\frac{1}{2}}q( (\phi^+)^2 + (\phi^-)^2) ) ^2.
\end{align}
We see that Equation \eqref{eq: decoLor} acts to decohere the quantum system into the $|\phi\rangle$ basis by suppressing configurations away from $\phi^+ = \phi^-$ by an amount proportional to $D_0 \lambda^2$, where $\lambda$ characterizes the back-reaction on the quantum system. On the other hand, Equation \eqref{eq: diffLor} acts to suppress configurations away from their semi-classical equations of motion - found from varying $\frac{\delta \bar{W}_{CQ}}{\delta q}$ - by an amount also proportional to $D_0$. Note that this does not depend on the coupling strength so that in the regime where the back-reaction is small, one can maintain coherence without deviating too much from the expected classical equations of motion. \jono{This can be used to evaluate CQ path integrals by working perturbatively in the back-reaction coupling. We show how this can be done in Appendix \ref{sec: peturbative}. }

Here, when the decoherence-vs-diffusion trade-off is saturated, the $\phi^+\phi^-$ coupling terms cancel. This is the term which corresponds to the {\it jump-jump} term in Lindbladian evolution. This makes the total action relatively simple, namely
\begin{align}\label{eq: two scalar fields}
     \mathcal{I}( \xql,  \xqr,  \q,t_i,t_f) &=    \int_{t_i}^{t_f} dx \bigg[   i\mathcal{L}_Q^+(\q,x)  - i\mathcal{L}_Q^-(\q,x)      
        -\frac{1}{2}\frac{ \delta  \Delta W_{CQ}}{ \delta q(x)} D_0 \frac{\delta \Delta W_{CQ}}{ \delta q(x)} 
         - \frac{1}{2}\frac{\delta \bar{W}_{CQ} }{ \delta q} D_{2}^{-1}(\q,x) \frac{\delta \bar{W}_{CQ} }{ \delta q(x)}
    \bigg]\nonumber\\
      &   = \int_{t_i}^{t_f} dx \bigg[   i\mathcal{L}_Q^+(\q,x)  - i\mathcal{L}_Q^-(\q,x)      
        -\frac{1}{4D_2}\left(\Box\q+\q\lambda(\phi^+)^2\right)^2
         -\frac{1}{4D_2}\left(\Box\q+\q\lambda(\phi^-)^2\right)^2
         \bigg]
\end{align}
in the $m=0$ case. Note that $i\mathcal{L}_Q^\pm(\q,x)$ includes the coupling $\pm i\frac{\lambda}{2}q^2\phi^{\pm}$. \new{Here we see that if $D_2$ is large, then paths away from the deterministic solutions $\Box\q+\q\lambda(\phi^\pm)^2=0$  will no longer be suppressed, while if $D_2$ is small (i.e. $D_0$ is large), the decoherence term will be enhanced. This is the essence of the decoherence-vs-diffusion trade-off.}

\noindent \textbf{Diffeomorphism invariant CQ gravity.} Let us now comment on some of the consequences of classical-quantum theories of gravity. The goal is to try and construct a theory of covariant classical-quantum dynamics that approximates Einstein's equations. 
\jono{We will find that we can construct a theory which is manifestly diffeomorphism invariant. We believe it's likely to be different to the theory presented in \cite{Oppenheim:2018igd} using master equation methods. In particular, we do not yet know if \cite{Oppenheim:2018igd} is diffeomorphism invariant owing to the constraint algebra being unsolved\cite{pqconstraints}. Nor do we know whether the two theories are equivalent, since one can generally only derive the path integral from the master equation, when the dynamics is at most quadratic in the momenta, which is not the case here.} 

Since in Equation \eqref{eq: fieldConfigurationSpace2} the paths away from $\frac{\delta }{ \delta q_i}(  \bar{W}_{CQ}[q, \xq^{\pm}])$ are exponentially suppressed by an amount depending on $D_2^{-1}$, the most likely path will be those for which 
\begin{equation}
\frac{\delta }{ \delta q_i}( \bar{W}_{CQ}[q, \xq^{\pm}]) \approx 0.
\end{equation}
To get a theory that agrees with Einstein's gravity on average, we could therefore try to take $W_{CQ}[g,\phi]$ to be the sum of the Einstein Hilbert action $S_{EH}[g] = \frac{1}{16 \pi G_N} \int \sqrt{g} R$, and a matter action $S_{m}[g, \phi]$ including a cosmological constant. The path integral is over physically distinct geometries, which we could enforce via the Fadeev-Popov procedure. In the case where $W_{CQ} = S_{EH} + S_{m}$ we have
\begin{equation}\label{eq:KGaction}
 \frac{\delta }{ \delta g_{\mu \nu}}( W_{CQ} [g, \phi])  = - \frac{\sqrt{-g}}{16 \pi G_N}( G^{\mu \nu} +\Lambda g^{\mu\nu}-  8 \pi G_N T^{ \mu \nu }),
\end{equation}
Thus, paths would be exponentially suppressed away from (a $\pm$ branch average of) Einstein's equations. Explicitly, taking the classical degree of freedom to be $g_{\mu \nu}$, the decoherence part of the CQ interaction in Equation \eqref{eq: fieldConfigurationSpace2} is given by 
\begin{equation}\label{eq: deco}
     \frac{\delta \Delta W_{CQ}}{\delta g_{\mu \nu}}D_{0,\mu \nu\rho \sigma}  \frac{\delta \Delta W_{CQ}}{\delta g_{\rho\sigma}} 
    =\det(-g)\frac{1}{4}( T^{\mu \nu + } - T^{\mu \nu - }) D_{0,\mu \nu\rho \sigma}(T^{\rho \sigma+ } - T^{\rho \sigma- }),
   \end{equation}
   whilst, assuming $4D_0 = D_2^{-1}$ the diffusion part takes the form 
   \begin{equation} \label{eq: diff}
     \frac{\delta \bar{W}_{CQ}}{\delta g_{\mu \nu}}D^{-1}_{2,\mu \nu\rho \sigma}  \frac{\delta \bar{W}_{CQ}}{\delta g_{\rho \sigma}}  = \frac{1}{64\pi^2 G_N^2} \det(-g) \times \
     (G^{\mu \nu}+\bar{\Lambda} g^{\mu\nu} -  8 \pi G_N \bar{T}^{ \mu \nu  }) D_{0,\mu \nu\rho \sigma}(G^{\rho \sigma} +\bar{\Lambda} g^{\mu\nu} -  8 \pi G_N \bar{T}^{ \rho \sigma  }).
\end{equation}
Because the decoherence-diffusion trade-off is saturated, the dynamics take the form of Equation \eqref{eq: positiveCQ}; thus, the dynamics are completely positive, and the quantum state of the fields remains pure conditioned on the metric. The interaction is fully characterized by the tensor density $D_{0,\mu \nu\rho \sigma}$. There are two possible demands one could make on this tensor. The first would be to require that it be a positive semi-definite matrix in the sense that $v^{\mu \nu} D_{0,\mu \nu\rho \sigma} v^{\rho \sigma} \geq 0$ for any matrix $v^{\rho \sigma}$. This would ensure that the dynamics are completely positive and normalizable on any initial state, and classical paths which are close to Einstein's equations are more probable. Constructing diffeomorphism invariant classical-quantum theories of gravity then amounts to trying to find a tensor  $D_0^{\mu \nu\rho \sigma}$ which gives rise to a path integral which defines completely-positive dynamics.

To meet this demand, the simplest thing one can try is to take $D_{0,\mu \nu \rho \sigma} = D_0 g^{-1/2} g_{\mu \nu} g_{\rho \sigma}$, in which case one finds a diffeomorphism invariant CQ theory of gravity in which paths deviating from the \textit{trace} of Einstein's equations are suppressed. Moreover, according to Equation \eqref{eq: deco} the quantum state decoheres into eigenstates of the trace of the stress-energy tensor. In the Newtonian limit, where the trace of the stress-energy tensor is dominated by its mass term, this acts to decohere the quantum state into mass eigenstates. This is related to the amplification mechanism used in spontaneous collapse models \cite{Pearle:1988uh,PhysRevA.42.78,BassiCollapse,PhysRevD.34.470, GisinCollapse, relcollapsePearle, 2006RelCollTum}, but here the decoherence mechanism is non-Markovian and arises as a consequence of treating the gravitational field classically and imposing diffeomorphism invariance on the CQ action. Furthermore, although the quantum state decoheres, it remains pure if we condition on the classical trajectory.

This is sufficient to demonstrate that a diffeomorphism invariant CQ theory of gravity is possible. The challenge in constructing a complete theory is to obtain the transverse parts of the Einstein equation, which are the constraints, whilst still ensuring the path integral over classical geometries remains negative definite so that the path integral converges. 

A general form for the diffusion matrix is to take it to be proportional to the generalized Wheeler-deWitt metric in $3+1$ dimensions
\begin{align}
D_{0,\mu \nu \rho \sigma}=\frac{1}{8 D_\gamma} (-g)^{-1 / 2}\left(g_{\mu\rho} g_{\nu \sigma}+g_{\mu \sigma} g_{\nu \rho}-2\beta g_{\mu\nu} g_{\rho \sigma}\right)
\end{align}
\jono{with  $D_\gamma:=D_2G_N^2/c^3$ the dimensionless coupling constant of the theory, and $D_2$ being the diffusion coefficient which was given units of ${kg^2 s m^{-3}}$ in \cite{dec_Vs_diff}. It can also be convenient to give both $D_2$ and $D_0$ units of $G_N/c^3$, but since we here saturate the trade-off between $D_2$ and $D_0$ we are left with only one of them.} 
While $D_{0,\mu \nu \rho \sigma}$ is not positive semi-definite,  we only require it to be positive semi-definite on physical, local degrees of freedom. \new{We have since shown that this is the case, by showing that the two-point function for the scalar mode\cite{grudka2023renorm} and tensor mode\cite{oppenheim2025tensormode} are positive semi definite kernels. It then follows that deviations away from the dynamical equations of general relativity are suppressed.} \cite{grudka2023renorm,UCLNormPI}. 
Furthermore, normalisation of the dynamics is tenable in part because the negative eigenvalues correspond to non-dynamical components of the path integral. 

One could conceive of other geometric terms, for example, one could consider various powers of the determinant of $G_{\mu\nu}$ and $T_{\mu\nu}$. On the other hand, if one were to choose a $D_{0,\mu \nu \rho \sigma}$, which was not purely geometric, it either introduces a preferred background or must be made dynamical. The former is more suggestive of an effective theory in which one obtains a classical metric by adding in decoherence or tracing out degrees of freedom in some reference frame. In the latter case, one should add terms proportional to $D_{0,\mu\nu\rho\sigma} g^{\mu\nu}g^{\rho\sigma}$ and $D_{0,\mu\nu\rho\sigma}g^{\mu\sigma}g^{\nu\rho}$ into the classical part of the action. One then needs to ensure that such terms are not in conflict with experiment.

It is also possible to consider a $D_{0,\mu \nu \rho \sigma}(x,x')$ which is a positive-definite kernel in space-time coordinates $x,x'$ in which case one has stochastic processes which are correlated in space-time, and the CQ interaction terms take the form
\begin{equation}\label{eq: decoNonlocal}
   -\frac{1}{2}\int d^4x d^4x' \ \frac{\delta \Delta W_{CQ}}{\delta g_{\mu \nu}(x)}D_{0,\mu \nu\rho\sigma} (x,x') \frac{\delta \Delta W_{CQ}}{\delta g_{\mu \nu}(x')} 
  -\frac{1}{2}\int dx dx' \ \frac{\delta \bar{W}_{CQ}}{\delta g_{\mu \nu}(x)}D^{-1}_{2,\mu \nu \rho \sigma}(x,x')  \frac{\delta \bar{W}_{CQ}}{\delta g_{\mu \nu}(x')} .
   \end{equation}
   Since these noise kernels have spatial correlations, Lorentz-invariance implies that they must also have temporal correlations and so they represent non-Markovian dynamics. This is suggestive of an effective CQ-theory rather than fundamental ones since non-Markovianity implies the existence of a hidden memory.
However, such a theory may have an advantage in terms of suppressing heating and diffusion.

\noindent \textbf{No mediation of entanglement by classical fields.}
\new{The proposed experiments of \cite{bose2017spin,marletto2017gravitationally} are based on the fact that two systems interacting via the gravitational field will not become entangled if the gravitational field is classical. The argument was that  local operations and classical communication (LOCC), cannot generate entanglement; a classical interaction can be thought of as a form of classical communication. This provides an experimental basis to test the quantum nature of spacetime~\cite{bose2017spin}. However, in the non-relativistic limit, the interaction between two small masses is dominated by the Newtonian interaction which directly couples the position of the masses and is non-dynamical (being a constraint of general relativity). Thus there has been considerable debate on whether a classical Newtonian interaction can generate entanglement~\cite{danielson2022gravitationally,husain2022dynamics,ma2022limits,gollapudi2025state,martin2023gravity,trillo2025diosi, marletto2025koopmanian}. }

\new{We will now see that any CQ theory which is local in the quantum fields, and action of the form of Eq. \ref{eq: positiveCQ} cannot generate entanglement via the classical field. It is noteworthy, that we do not need to assume that the classical interaction is local. For ease of presentation, we will first take all the $c^{\gamma}=0$ (it is straightforward to show that if they are non-zero, they just produce additional local decoherence which cannot generate entanglement). The first key step is one we've already discussed.  If $c^{\gamma}=0$, then an action of the form of Eq. \ref{eq: positiveCQ} preserves the purity of the quantum state conditioned on the trajectory $\Phi_t$ of the classical field $\Phi$. In particular, the action of the CQ dynamics on the quantum state for a given trajectory $\Phi_t$ is proportional to a single operator $L(\Phi_t)$.  For any initial pure quantum state $|\Psi_i\rangle$ of the matter distribution and initial configuration $\Phi_i$ of the Newtonian potential, we write the initial CQ density matrix entries as $\rho(\Phi_i,m^+_i,m^-_i, t_i)=\langle m^+|\Psi_i\rangle\langle \Psi_i|m^-\rangle \delta(\Phi-\Phi_i)$ and final state given by
\begin{align}
    \rho(\Phi_f,m_f^+,m_f^-,t_f)  = &
    \int \mathcal{D}\Phi_t \mathcal{D} m^+ \mathcal{D} m^- \N e^{\mathcal{I}[\Phi,m^+,m^-,t_i,t_f]}   \rho(\Phi_i,m^+_i,m^-_i, t_i).\nonumber\\
    =&  \int \mathcal{D}\Phi_t \mathcal{D} m^+ \mathcal{D} m^- \N e^{-I_C[\Phi]}e^{\mathcal{I}_{CQ}[\Phi,m^+,t_i,t_f] } \langle m^+|\Psi_i\rangle\langle \Psi_i|m^-\rangle \delta(\Phi-\Phi_i)e^{\mathcal{I}_{CQ}[\Phi,m^-,t_i,t_f]} \nonumber\\
    = & \int \mathcal{D}\Phi_t  \langle m^+|L(\Phi_t)|\Psi_i\rangle\langle \Psi_i| L^\dagger(\Phi_t)|m^-\rangle \delta(\Phi-\Phi_f).\label{eq: kruasPI}
\end{align}
This is a special case of Equation \eqref{eq: CPmap}, with the Kraus operators $L(\Phi_t)$ labeled by a continuous index, given by the trajectories of the classical field. The total evolution is thus an incoherent sum (an integral) over all possible transitions of the quantum state.}

\new{We next want to make more precise the requirement that the degree of freedom $m(x)$ that we want to entangle, interacts primarily via the classical field. This is required in any experiment since the quantum fields can generate entanglement and need to either be screened or otherwise made small. We will do so by demanding that the quantum interaction is {\it localised}. For two regions $A$ and $B$, and for any fixed $\Phi$, we say a CQ action is {\it localised}, if it is of the form 
\begin{align}
    \mathcal{I}_{CQ}[\Phi,m^\pm,t_i,t_f] \approx  \mathcal{I}_{A}[\Phi,m_A^{\pm}t_i,t_f]
    +\mathcal{I}_{B}[\Phi,m_B^{\pm},t_i,t_f]
    \label{eq: locality}
\end{align}
 with  $\mathcal{I}_{A}[\Phi,m_A^\pm, t_i,t_f]$ only depending on the matter field $m(x)$ in region $A$  and  $\mathcal{I}_{B}[\Phi,m_B^\pm,t_i,t_f]$ only depending on the matter field $m(x)$ in region $B$. In other words, the quantum fields don't couple region $A$ with region $B$. If such a condition is met,  the operators corresponding to $\mathcal{I}_A$ and $\mathcal{I}_B$ act on separate Hilbert space factors, and commute, allowing the exponential operators to be separated $e^{\mathcal{I}_{CQ}}\approx e^{\mathcal{I}_{A}}e^{\mathcal{I}_{B}}$, and thus we can see from Eq. \eqref{eq: kruasPI} that this implies  $L(\Phi_t)\approx L_A(\Phi_t)\otimes L_B(\Phi_t)$.
 This enables us  to write the dynamics as the map
 \begin{align}
\varrho(\Phi_f,t_f)=   \int  \mathcal{D}\Phi_t   L_A(\Phi_t)\otimes L_B(\Phi_t)\varrho(\Phi_i,t_i)L^\dagger_A(\Phi_t)\otimes L^\dagger_B(\Phi_t).
\label{eq: productKraus}
 \end{align}
Such a map cannot create entanglement between region $A$ and $B$, since acting it on a product state $|\Psi_A\rangle\otimes|\psi_B\rangle$, produces a statistical mixture of product states, which is a separable state. The path integral approach provides therefore provides a direct proof that local CQ dynamics cannot generate entanglement. 
}

\new{Let us now discuss which CQ-models satisfy the localised assumption of Eq. \eqref{eq: locality}. It requires (i) that the Lindbladian have a  sufficiently local noise kernel, and (ii) that any quantum interactions between the relevant degrees of freedom in $A$ and $B$ be negligible or screened with some electrostatic shielding. These are both necessary condition -- Lindbladians with non-local noise kernels can generate entanglement~\cite{oppenheim2009fundamental}, including those which implement Diosi-Penrose decoherence, or the Tilloy-Diosi model~\cite{tilloy2016sourcing, trillo2025diosi}. 
Since these non-local noise kernel's are suggestive of an effective CQ-theory rather than a fundamental one, as discussed in relation to the kernels of \eqref{eq: decoNonlocal}, we should not be surprised that they can create entanglement. This is also natural since they can be viewed as an interaction with an entangled environment\cite{tilloy2016sourcing}.
We have already discussed why requirement (ii) is needed. Quantum interactions between $A$ and $B$ such as electromagnetism needs to be made negligible because these can create entanglement.}

\new{The local models discussed here, satisfy the locality condition of Eq. \eqref{eq: locality} whenever (ii) is satisfied. As a specific example, we consider the Newtonian limit of the gravity path integral from the previous section. We consider two mass densities $m_A(x)$ and $m_B(x)$ in each of the regions, interacting via a classical Newtonian potential $\Phi$. The Newtonian limit of our gravity action  is given by~\cite{layton2023weak}
\begin{equation}
\label{eq: PQG-actionNewton2}
\begin{split}
    \mathcal{I}_{CQ}[\Phi,m^+,m^-,t_i,t_f] = \int_{t_i}^{t_f} d^4x\, &\bigg[i\big(\mathcal{L}_{Q}[m^+]-\mathcal{V}_I[\Phi,m^+]-\mathcal{L}_{Q}[m^-]+\mathcal{V}_I[\Phi,m^-]\big)\\& -\frac{1}{8\tilde{D}_2}\big( m^+(x) - m^-(x)\big)^2  - \frac{1}{2\tilde{D}_2} \left( \frac{\nabla^2 \Phi}{4 \pi G } - \bar{m}(x)\right)^2 \bigg],
\end{split}
\end{equation}
where $\bar{m}(x)=\frac{1}{2}\big(m^+(x)+m^-(x)\big)$ and $\mathcal{V}_I[\Phi,m^\pm] \approx -2\Phi m^\pm$, coming from expanding the $\sqrt{-g}$ in the matter action. $\mathcal{L}_{Q}[m^\pm]$ is the matter action in Minkowski space.  We can see from this action that $\Phi$ satisfies Poisson's equation, sourced by both the bra and ket mass density, with the size of deviations from this controlled by $\tilde{D}_2$. }

\new{Let us first consider the Lindbladian term. It is ultra-local, so this part of the action taken in isolation can be split into a part which acts in region $A$ and another part which acts in region $B$
\begin{align}
    \int d^4x \big( m^+(x) - m^-(x)\big)^2 
    \approx
    \int d^4x \big( m_A^+(x) - m_A^-(x)\big)^2 +\int d^4x \big( m_B^+(x) - m_B^-(x)\big)^2\,.
\end{align}
By way of comparison, the correlated Diosi-Penrose noise kernel 
$\int d^3xd^3y\frac{1}{|x-y|}\big( m^+(x) - m^-(x)\big)\big( m^+(y) - m^-(y)\big) $ cannot be split into a sum of an $A$ part and a $B$ part, unless $m_A(x)m_B(y)/|x-y|$ is negligible.}

\new{The remaining part of the matter action are the bra and ket actions $\int d^4 x \bigg[\pm i\mathcal{L}_{Q}[m^\pm]\mp 2i \Phi m^\pm- \frac{1}{2\tilde{D}_2}  \frac{\nabla^2 \Phi}{4 \pi G }  m^\pm\bigg]$. The last two terms don't have any temporal or spatial derivatives of $m^\pm(x)$ so the question of whether the action has the form of Eq. \eqref{eq: locality} hinges on the form of $\mathcal{L}_{Q}[m^\pm]$. If gravity is the dominant force, the matter action is dominated by the rest mass density $\mathcal{L}_{Q}[m^\pm]=m^\pm(x)$  and we can neglect  kinetic terms. This action for the matter field is local in the sense of Eq. \eqref{eq: locality}, and thus cannot generate entanglement between separated regions.
 On the other hand, if we have other interactions between region $A$ and $B$, then Eq. \eqref{eq: locality} need not be satisfied, and these other interactions can create entanglement. For example, if we have a scalar field with  $\mathcal{L}_{Q}[\phi^\pm]=\frac{1}{2}(\partial\phi)^2-\frac{1}{2}m^2\phi^2$, then the  $(\partial\phi)^2$ term means that Eq. \eqref{eq: locality} doesn't hold. An easy way to check this is to add a source for the quantum field restricted to $A$ and $B$ $J_{A\pm}(x)\phi^\pm(x)+J_{B\pm}(x)\phi^\pm(x)$ into the action and perform the Gaussian integral over $\phi$, which induces couplings $\frac{J_A(x)J_B(y)}{|x-y|}$ into the effective action. }

\new{By way of comparison, if we treated the Newtonian potential as a quantum field, with action
\begin{equation}
\label{eq: QG-actionNewton}
\begin{split}
    \mathcal{I}_{Q}[\Phi^+,\Phi^-,m^+,m^-,t_i,t_f] = \int_{t_i}^{t_f} d^4x\, &i\bigg[\mathcal{L}_{Q}[m^+]-\mathcal{V}_I[\Phi^+,m^+]-\mathcal{L}_{Q}[m^-]+\mathcal{V}_I[\Phi^-,m^-]\\&+\frac{1}{4\pi G}(\nabla\Phi^+)^2 -\frac{1}{4\pi G}(\nabla\Phi^-)^2\bigg],
\end{split}
\end{equation}
then we would not be able to write the evolution in terms of product Kraus operators with an index given in terms of a single classical field trajectory $\Phi_t$ as in Eq. \eqref{eq: productKraus}. The fact that we can write the evolution as a convex mixture of products, plus the short range of the quantum interaction and Lindbladian is what prohibits the generation of entanglement. The classical interaction itself need not be local.  In contrast, given the fully quantum action of Eq. \eqref{eq: QG-actionNewton} and source terms $m^\pm_A(x)\Phi(x)+m^\pm_B(x)\Phi^\pm(x)$, one can perform the Gaussian integral over the bra and ket Newtonian potentials $\Phi^+$ and $\Phi^-$, to get an interaction which clearly can create entanglement between regions $A$ and $B$ as one has terms $\pm iG\frac{m_A^\pm(x)m^\pm_B(y)}{|x-y|}$ in the reduced action~\cite{christodoulou2023locally}.}

\noindent \textbf{Discussion.} In this work, we have introduced a general path integral for classical-quantum dynamics, given by Equation \eqref{eq: positiveCQ}, which opens up the way to study classical degrees of freedom coupled to quantum ones via path integral methods. This provides an approach to study covariant theories of classical fields coupled to quantum ones. We have given an explicit example of a Lorentz invariant CQ theory and  \new{applied it} to classical-quantum theories of gravity.  

In particular, we have arrived at a diffeomorphism invariant theory of CQ general relativity - summarized by Equations \eqref{eq: deco}, \eqref{eq: diff} - which acts to suppress paths that deviate from the trace of Einstein's equations, whilst simultaneously decohering the quantum system according to the trace of the stress-energy tensor. This provides a first example of diffeomorphism invariant classical-quantum dynamics and, more generally, is a first example of diffeomorphism invariant collapse dynamics \cite{Pearle:1988uh,PhysRevA.42.78,BassiCollapse,PhysRevD.34.470, GisinCollapse, relcollapsePearle, 2006RelCollTum}, where the loss of coherence is a derived consequence of the interaction of a quantum system with a classical dynamical variable. We have also proposed a diffeomorphism invariant theory that reproduces all of Einstein's equations as a limiting case.  Since $D_{2,\mu\nu\rho\sigma}$ is not positive semi-definite, we have not proven that the dynamics is normalizable, and suppresses paths away from Einstein's equation, but \new{this has since been shown in \cite{UCLNormPI,oppenheim2025tensormode}}.

We have here given a general construction by which one can write down CQ path integrals that uphold space-time and gauge symmetries. It would be worthwhile to explore this further with concrete examples. For classical-quantum gauge theories, which could be useful in an effective theory of light-matter interactions when there is classical back-reaction, the killing form provides a natural choice for $D_0$ since for a compact lie group, the killing form is positive semi-definite \cite{Fuchs:1997jv}. 

\jono{The theory and formalism presented here has a number of applications. Let us first note some of those which have been made since this work appeared on the arxiv\cite{oppenheim2023covariant-arxiv}.
We have since shown in \cite{layton2023weak,oppenheim2024diffeomorphism} that the non-relativistic weak field limit of the gravitational path integral, reproduces the weak field limit of the theory derived using the master equation method of \cite{Oppenheim:2018igd} and that deriving using a measurement and feedback approach \cite{tilloy2016sourcing}. However, there is an important difference. While the non-relativistic but local theories of \cite{tilloy2016sourcing,layton2023weak} are ruled out by experiment via the decoherence-diffusion trade-off, due to having an IR divergence\cite{dec_Vs_diff}, we find that the relativistic theory presented here is not \cite{grudka2023renorm}. } 

\jono{
In \cite{oppenheim2023covariant-arxiv}, we noted that that since the propagator scaled like $1/p^4$ the theory could be renormalisable. We have since shown that the pure gravity path integral presented here is formally renormalisable without having tachyons or ghosts\cite{grudka2023renorm}. A full proof of renormalizability would require showing that the pole prescription which results in the theory being renormalizable also retains the property that it is completely positive. This was shown for the scalar modes.  Though effective theories can be non-renormalizable, the renormalizability of CQ dynamics in the gravitational degrees of freedom has important foundational consequences since the prime motivation for believing that gravity may not be a quantum field, is that it reflects the curvature of space-time. If that is the  case, then the description of gravity in terms of the metric ought to be a fundamental description which should not break down at some energy scale. In contrast, perturbative quantum gravity is not renormalizable in $3+1$ dimensions, so it is unclear if the geometrical description of gravity would hold in the quantum theory.
}

For the matter degrees of freedom, the gravitational action of Equation's \eqref{eq: deco} and \eqref{eq: diff} are not power counting renormalizable due to the terms which are quadratic in the stress-energy tensor, though, as noted in \cite{Baidya:2017eho, Avinash:2019qga}, one must be careful with power counting renormalization when considering the density matrix path integral. \jono{The additional higher order terms which are of most interest,  are terms such as the mass density squared, which for the scalar field go as $(\phi^\pm)^4$. These serve to decohere the quantum system into mass distributions.  On the other hand, the non-renormalizable terms with negative mass dimension such as $(\partial\phi^\pm)^4$ act to decohere the state into kinetic energy distributions, and at low energy are suppressed by higher derivatives.  From both an effective field theory, and in terms of backreacting on the gravitational field, such terms are not relevant at low energy.}

\jono{From an observational point of view, perhaps the most significant application of the path integral introduced here, is that it enables the calculation of the two-point function of the gravitational field. This has since been undertaken in \cite{grudka2023renorm} for the scalar mode.
The spectral density of the scalar mode for $\nabla\phi$,  measured at frequency $\omega$ for a ''mod-squared Feynman'' pole-prescription was found to be\cite{grudka2023renorm}
\begin{align}
\label{eq:thespectraldensity}
     S_{aa}(\vec{x},\vec{x}';\omega)|_{\omega_0^2, \omega\rightarrow 0}=&\frac{D_\gamma}{4\pi \omega_0 |x-x'|}\left(\omega^2 \sin(\omega|x-x'|)+\omega_0^2\cos(\omega|x-x'|)\right)
     \nonumber\\
     \approx&\frac{D_\gamma}{4\pi}\left(\frac{\omega^3}{\omega_0^2}+\frac{1}{|x-x'|}\right)
\end{align}
where we take $\omega_0$ to be some minimal observable frequency such as that given by the inverse of the Hubble time. At low frequency, we see that there are local stochastic fluctuations, and much stronger but longer range fluctuations which are likely not observable in table top experiments due to being uniform over large distances. \new{The two-point function for a ''mod-squared retarded'' pole prescription is found in \cite{oppenheim2025stochasticwave} and shown to be appropriate for the stochastic Klein-Gordon equation. Understanding the relationship between the Newtonian potential and this scalar mode requires a deeper understanding of coordinate freedoms and diffeomorphism invariance. } This provides a path towards strong experimental constraints on the theory, through the decoherence vs diffusion trade-off \cite{dec_Vs_diff}. Precision acceleration experiments set an upper bound on $D_\gamma$, while the trade-off implies that interference experiments put a lower found on $D_\gamma$, thus constraining the dimensionless coupling constant of the theory from both sides, and possibly falsifying the theory. Current experimental bounds put $10^{-64} \geq D_\gamma\geq 10^{-54}$,  a gap which may seem large, but which could be closed in the near-term via interference experiments with heavy atoms in narrowly peaked superpositions \cite{grudka2023renorm}.
}

We have here approached CQ dynamics from a bottom-up approach: starting from the description of a system in terms of classical and quantum variables, we have written down a description for the dynamics which leads to consistent evolution. It would be interesting to arrive at classical-quantum theories from a top-down approach. That is, starting from a quantum-quantum system, we should be able to arrive at an effective classical-quantum description. \jono{We have since shown with Isaac Layton, that there is a parameter range of the classical-quantum dynamics presented here, which arises from a ''classical-quantum limit'' of two quantum systems \cite{layton2024classical}. This occurs via a decoherence mechanism on one of the systems and is closely related to the quantum to classical transition \cite{zurek1982environment,paz1993environment,zurek2006decoherence,Brun_2002}. Such an approach would be useful as an effective theory of semi-classical gravitational physics when back-reaction is involved, for example, in inflationary cosmology or during black-hole evaporation.  As a first step, one would like to extend the result of \cite{layton2024classical} to the field theoretic case, whose natural setting is the path integral formulation described here. Since the pure gravity path integral is renormalizable, one might hope that the effectively classical-quantum theory retains this feature. Thus although perturbative quantum gravity in $3+1$ dimensions might not be renormalisable, it might be in the limit that the system becomes classical.} Aside from CQ gravity models, the covariant path integral has subsequently been used to develop an alternative effective theory of wavefunction collapse by coupling a classical scalar field to perturbative quantum gravity \cite{Weller-Davies:2024zcb}. This could also be related to quantum gravity subject to a decohering environment.

\jono{
We have presented a simple model of Lorentz invariant classical-quantum field theory.
In Appendix \ref{sec: peturbative} we initiate the study of two interacting Lorentz invariant scalar fields, one classical, one quantum, by demonstrating the use of perturbation theory to compute the partition function, as well as methods for computing the normalisation in \ref{app: normalizing}. A slightly simpler model in which the classical-quantum interaction is linear in the classical field $q\phi$ has since been presented in \cite{grudka2023renorm}. Recently Carney and Matsumura computed the scattering cross-section of this model~\cite{carney2024classicalquantumscattering}, and confirmed both the Lorentz invariance of the result, and the non-Markovian nature of the dynamics when the classical field is integrated out. They also found that if one treats planets as point particles, and assuming Markovian dynamics, that there are order one corrections to the Rutherford result. This appears to be related to the fact that diffusion in the classical system can induce secondary decoherence in the quantum system \cite{tilloy2017principle}, which can result in anomalous heating \cite{bps,gross1984quantum}.  This secondary effect, and suggestions for suppressing it were discussed in \cite{Oppenheim:2018igd}, including adding higher order terms in the classical field as are found in general relativity, as well as friction terms, or field dependent diffusion co-efficients. More generally, it is hoped that the methods presented here provide a sufficient template to explore this and other issues in a variety of models which respect spacetime and other symmetries. 
}

\section*{Acknowledgements}
We would like to thank Maite Arcos, Joan Camps, Andrea Russo, Carlo Sparaciari, Barbara \v{S}oda, and Edward Witten for valuable discussions. We are grateful to Isaac Layton and Emanuele Panella for clarifying the role of the normalization in the path  integral. JO is supported by an EPSRC Established Career Fellowship, and a Royal Society Wolfson Merit Award. This research was supported by the National Science Foundation under Grant No. NSF PHY11-25915 and by the Simons Foundation {\it It from Qubit} Network.  Research at Perimeter Institute is supported in part by the Government of Canada through the Department of Innovation, Science and Economic Development Canada and by the Province of Ontario through the Ministry of Economic Development, Job Creation and Trade.
\bibliography{refCQ}

\begin{thebibliography}{76}
\bibitem [{UCL()}]{UCLNormPI}%
  \BibitemOpen
  \href@noop {} {}\bibinfo {note} {There are subtleties regarding the
  normalisation which have since been clarified with Emanuele Panella and Isaac
  Layton (Manuscript in preparation, 2023)}\BibitemShut {NoStop}%
\end{thebibliography}

\appendix
\onecolumngrid
\newpage

\section{Comparison of classical, quantum and classical-quantum path integrals}\label{sec: comparison}

The classical-quantum path integral generalizes the Feynman-Vernon path integral of open quantum systems and the stochastic path integral of classical systems. In \cite{UCLPILONG}, we compare and contrast these three different path integrals, and for convenience, we include Table \ref{tab: pathintegrals} below.\begin{table}[H]

\caption{\label{tab: pathintegrals}A table representing the classical, quantum, and classical-quantum path integrals.}
\small
\begin{subtable}{\linewidth}
\begin{tblr}{
  colspec = {X[2cm,c]X[c]},
  stretch = 0,
  rowsep = 4pt,
  hlines = {black, 0.5pt},
  vlines = {black, 0.5pt},
}
 & Classical stochastic  \\

    \parbox{\linewidth}{Path integral} & $p(q, p,t_f) = \int \mathcal{D} q \mathcal{D} p \ e^{iS_C[q,p]} \delta( \dot{q} - \frac{\partial H}{\partial p}) p(q,p,t_i) $  \\ 

\parbox{\linewidth}{Action} & $iS_C =-\int_{t_i}^{t_f} dt \frac{1}{2} \ (\frac{\partial H}{\partial q} + \dot{p}) D_2^{-1} (\frac{\partial H}{\partial q}  + \dot{p}) $  \\ 

\parbox{\linewidth}{CP condition} &     $D^{-1}_2$ a positive (semi-definite) matrix, $D^{-1}_2 \succeq 0 $  \\

\end{tblr}

\caption{The path integral for continuous, stochastic phase space classical dynamics \cite{Onsager1953Fluctuations, freidlin1998random,Weber_2017, Kleinert}. One sums over all classical configurations $(q,p)$ with a weighting according to the difference between the classical path and its expected force $-\frac{\partial H}{\partial q}$, by an amount characterized by the diffusion matrix $D_{2}$. In the case where the force is determined by a Lagrangian $L_c$, the action $S_C$ describes suppression of paths away from the Euler-Lagrange equations $iS_C =-\int_{t_i}^{t_f} dt \frac{1}{2} (\frac{\delta L_c }{\delta q_i}) (D_2^{-1})^{ij} (\frac{\delta L_c }{\delta q_j}) $, by an amount determined by the diffusion coefficient $D_2$. The most general form of classical path integral can be found in \cite{Weber_2017, Kleinert, UCLPILONG}  } 
\end{subtable}
\vspace{0.1cm}

\begin{subtable}{\linewidth}
\begin{tblr}{
  colspec = {X[2cm,c]X[c]},
  stretch = 0,
  rowsep = 4pt,
  hlines = {black, 0.5pt},
  vlines = {black, 0.5pt},
}
 & Quantum    \\

    \parbox{\linewidth}{Path integral} & $\rho(\phi^{\pm}, t_f) = \int \mathcal{D} \phi^{\pm} \ e^{iS[\phi^+] - iS[\phi^- ] + iS_{FV}[\phi^+, \phi^-] } \rho(\phi^{\pm}, t_i)$   \\ 

\parbox{\linewidth}{Action} & $\displaystyle 
\begin{aligned}
& S[\xq]  = \int_{t_i}^{t_f} dt \big( \frac{1}{2}\dot{\xq}^2 + V(\xq) \big), \ \ \ iS_{FV} =  \int_{t_i}^{t_f} dt \big ( D^{\alpha \beta}_0  \lin_{\alpha}^{\rind} \lin_{\beta}^{* \lind} - \frac{1}{2} D^{\alpha \beta}_0 (L^{* \lind}_{\beta}\lin_{\alpha}^{\lind} + L^{*\rind}_{\beta}\lin_{\alpha}^{\rind} ) \big) 
 \end{aligned}$ \\

\parbox{\linewidth}{CP condition} &   $D_0^{\alpha \beta}$ a positive (semi-definite) matrix, $D_0 \succeq 0$. \\

\end{tblr}

\caption{\label{tab: quantumPI}The path integral for a general autonomous quantum system, here taken to be $\phi$. The quantum path integral is doubled since it includes a path integral over both the bra and ket components of the density matrix, here represented using the $\pm$ notation. In the absence of the Feynman Vernon term $S_{FV}$ \cite{FeynmanVernon}, the path integral represents a quantum system evolving unitarily with an action $S[\phi]$. When the Feynman Vernon action $S_{FV}$ is included, the path integral describes the path integral for dynamics undergoing Lindbladian evolution \cite{Lindblad:1975ef, GSKL} with Lindblad operators $L_\alpha(\phi)$. Because of the $\pm$ cross terms, the path integral no longer preserves the purity of the quantum state, and there will generally be decoherence by an amount determined by $D_0$. As an example, taking  $\lin^{\pm}=\phi^\pm (x)$ a local field, and  
 $D_0^{\alpha \beta}=D_0$ results in a Feynman-Vernon term
$iS_{FV} =  - \frac{1}{2}D_0 \int_{t_i}^{t_f} dt dx \big (\phi^\lind(x)-\phi^\rind(x)\big) ^2$ which decoheres the state in the $\phi(x)$ basis, since off-diagonal terms in the density matrix, where $\phi^\rind(x)$ is different to $\phi^\lind(x)$, are suppressed.
}
\end{subtable}

\vspace{0.3cm}

\begin{subtable}{\linewidth}
\begin{tblr}{
  colspec = {X[2cm,c]X[c]},
  stretch = 0,
  rowsep = 4pt,
  hlines = {black, 0.5pt},
  vlines = {black, 0.5pt},
}
 & Classical-quantum   \\

    \parbox{\linewidth}{Path integral} & $ \rho(q,p, \phi^{\pm}, t_f) = \int \mathcal{D} q  \mathcal{D} p \mathcal{D} \phi^{\pm} \ e^{iS_C[q,p] + iS[\phi^+] - iS[\phi^{-} ] + iS_{FV}[\phi^{\pm}]+ iS_{CQ}[q,p, \phi^{\pm}]  } \delta( \dot{q} - \frac{p}{m}) \rho(q,p,\phi^{\pm}, t_i)
    $\\ 

\parbox{\linewidth}{Action} & $\displaystyle 
\begin{aligned}
& iS_C[z] + iS_{CQ}[z,\phi^{\pm}] = -\frac{1}{2} \int_{t_i}^{t_f} dt \ D_2^{-1}\big( \frac{\partial H_c}{\partial q}+ \frac{1}{2} \frac{\partial V_I[q,\phi^+]}{\partial q } + \frac{1}{2} \frac{\partial V_I[q,\phi^-]}{\partial q } + \dot{p}\big)^2.
 \end{aligned}$ \\

\parbox{\linewidth}{CP condition} &   $D_0 \succeq 0, D_2 \succeq 0$ and $4D_2 \succeq  D_0^{-1}$ \\

\end{tblr}
\caption{The phase space path integral for continuous, autonomous classical-quantum dynamics. The path integral is a sum over all classical paths of the variables $\z$, as well as a sum over the doubled quantum degrees of freedom $\phi^{\pm}$. The action contains the purely quantum term from the quantum path integral in Table \ref{tab: quantumPI}, but also includes the term $iS_C+iS_{CQ}$. This suppresses paths away from the averaged drift, which is sourced by both purely classical terms described by the Hamiltonian $H_c$ and the back-reaction of the quantum systems on the classical ones, described by a classical-quantum interaction potential $V_I$. The most general form of classical-quantum path integral can be found in \cite{UCLPILONG}. In order for the dynamics to be completely positive, the decoherence-diffusion trade-off $4D_2 \succeq D_0^{-1}$ must be satisfied \cite{UCLPawula,dec_Vs_diff}, where $D_0^{-1}$ is the generalized inverse of $D_0$, which must be positive semi-definite. When the trade-off is saturated, the path integral preserves the purity of the quantum state, conditioned on the classical degree of freedom \cite{UCLHealing}.
}
\end{subtable}
\label{tab: MasterEquationTable} 
\end{table}
\section{Proof of positivity}\label{sec: proofofpos}
In this section, we prove the statement made in the main body that Equation \eqref{eq: positivelooking} defines completely positive CQ dynamics. To see this in detail, we can perform a short-time expansion of the full path integral, which we can always do since we assume the dynamics are time-local.

Let us first consider the case where the quantum state remains pure, so that $c^{\gamma}=0$ in Equation \eqref{eq: positivelooking}. Defining $t_f= t_0 + K  \delta t$, $t_i=t_0 + i \delta t$, and discretizing the path integral into steps of size $\delta t$ we have that 
\begin{equation}\label{eq: transitionAppendix}
    \cqstate_{i+1} = \int d\phi_i^+ d\phi^-_i dq_i ( e^{\mathcal{I}^+_{i+1,i}})( e^{\mathcal{I}^-_{i+1,i}})^* e^{-\mathcal{I}_{C,i+1,i}} \cqstate_i,
\end{equation}
where we use the shorthand $\cqstate_i = \cqstate(\phi^+_i,\phi^-_i,\q_i,t_i) $, $\mathcal{I}_{i+1,i} =\mathcal{I}[\phi_{i+1},\phi_i,\q_{i+1},\q_i]$ and $\mathcal{I}_{C,i+1,i} = \mathcal{I}_{C}[q_{i+1},q_i] $. 

More generally, we can allow for the case where the action contains higher time derivatives, in which case we have $\mathcal{I}[\phi_{i+k_q,}\dots,\phi_i,\q_{i+k_c},\dots,\q_i]$ and $\mathcal{I}_{C}[q_{i+k_c},q_i]$ with $k_c,k_q \geq 2$. In order to retain the usual composition law for the path integral, we must also let the state be described by increasingly higher derivative terms  $\cqstate(\phi^{\pm}_{i}, \dots, \frac{d^{k_q-1}\phi^{\pm}_{i}}{dt^{k_q-1}}, q_i,\frac{d^{k_c-1}\phi^{\pm}_{i}}{dt^{k_c-1}} ).$ \cite{Hawking:2001yt}. The final state then imposes boundary conditions on the components of the action, which contain higher derivative terms so that Equation \eqref{eq: transitionAppendix} is still well defined. 

With this in mind, we can take the trace with respect to an arbitrary vector
$|v(q)\rangle$, and for complete positivity, we need to show 
\begin{equation}\label{eq: poscond}
    \int d\phi_{i+1}^+ d\phi^-_{i+1} v_{i+1}^{+*} \cqstate_{i+1}v_{i+1}^- \geq 0 .
    \end{equation}
Denoting $\tilde{v}_{i+1,i}^+ =e^{\mathcal{I}^+_{i+1,i}} v_{i+1}^{+*} $, then inserting Equation \eqref{eq: transitionAppendix} into Equation \eqref{eq: poscond} we have
    \begin{equation}\label{eq: poscond2Append}
\int d\phi_{i+1}^+ d\phi^-_{i+1} d\phi_i^+ d\phi^-_i dq_i \tilde{v}_{i+1,i}^+ \tilde{v}_{i+1,i}^{-*} e^{-\mathcal{I}_{C,i+1,i}} \cqstate_i .
\end{equation}

Because the integral factorizes into $\pm$ conjugates, Equation \eqref{eq: poscond2Append} will always be positive. To see this explicitly, we first perform the $\phi^{\pm}_i$ integrals to obtain 
\begin{equation} \label{eq: firstint}
    c_{i+1} = \int d\phi_i^+ d\phi^-_i  \tilde{v}_{i+1,i}^+ \tilde{v}_{i+1,i}^{-*}\cqstate_ . \geq 0 ,
\end{equation}where we have used the positivity of the state CQ $\cqstate(\phi^+_i,\phi^-_i,\q_i,t_i) $. What remains is the integral 
\begin{equation}\label{eq: positivefinal}
    \int d\phi_{i+1}^+ d\phi^-_{i+1}  dq_i e^{-\mathcal{I}_{C,i+1,i}} c_{i+1} \geq 0,
\end{equation} 
which is positive since both $c_{i+1}$ and the exponential are both positive. In Equation \eqref{eq: positivefinal}, there is still a free $q_{i+1}$ variable which corresponds to the fact that positivity of the CQ state demands that the CQ dynamics keep quantum states positive conditioned on the classical degrees of freedom. We thus see that the state after applying the time-evolved state will also be positive. Hence the dynamics are positive. When we consider the dynamics as part of a larger system, we apply the identity map on the larger system, and the dynamics still factorize in this way  - we perform a delta function path integral on the auxiliary system, so Equation \eqref{eq: positiveCQ} defines completely positive dynamics. 

In the more general case, we can have non-zero $c^{\gamma}[q,x]$, and there is information loss since the dynamics can send pure states to mixed states. In this case, the only thing which changes is the definition of $\tilde{v}_{i+1,i}^+$ in Equation \eqref{eq: poscond}. In particular, in the general case, we must also expand out the terms involving $c_n^{\gamma}$ in the action of Equation \eqref{eq: positiveCQ}
\begin{equation}
     e^{\delta t \sum_{\gamma} c^{\gamma}_{i+1,i} (L^{+}_{\gamma})_{i+1,i} (L^{- }_{\gamma})^*_{i+1,i}} =  1+\delta t\sum_{\gamma} c^{\gamma}_{i+1,i} (L^{+}_{\gamma})_{i+1,i} (L^{-}_{\gamma})^*_{i+1,i}  +O(\delta t^2),
\end{equation}
where $c_{i+1,i}^{\gamma} = c^{\gamma}[q_{i+1},q_i]$, $L^+_{i+1,i} = L^+[\phi_{i+1}^+,\phi_i^+]$ and similarly for the $^-$ branch. 

With this in mind, the integrand of the path integral in Equation \eqref{eq: positiveCQ} factorizes according to Equation \eqref{eq: poscond2Append}, and the steps to prove complete positivity are exactly the same but with 
\begin{equation}
\tilde{v}_{i+1,i}^{\gamma +} 
= (1+\sqrt{\delta t c^{\gamma}_{i+1,i} } (L^{+}_{\gamma})_{i+1,i})e^{\mathcal{I}^+_{i+1,i}} v_{i+1}^{+*} , 
\end{equation}
from which the complete positivity of the dynamics follows from the same arguments outlined in Equation's \eqref{eq: poscond2Append} and \eqref{eq: firstint}, where we now also sum over $\gamma$. Note, though we need only work to first order in $\delta t$,  had we included them, the higher order $\delta t$ terms also factorize in the same way.

In the field-theoretic case, the total CQ action is 
 \begin{equation}\label{eq: positiveCQApp}
 \begin{split}
      \mathcal{I}( \xql,  \xqr,  \q,t_i,t_f) & =  \mathcal{I}_{CQ}(\q,\phi^+, t_i,t_f) + \mathcal{I}^*_{CQ}(\q,\phi^-, t_i,t_f)  - \mathcal{I}_C(\q, t_i,t_f)\\
      & + \int_{t_i}^{t_f} dx \sum_{ \gamma} c^{\gamma}(q,t,x)( L_{\gamma}[\phi^+](x)L^{*}_{\gamma}[\phi^-](x))
 \end{split}
\end{equation}
and we can repeat the argument for complete positivity, which again follows from the factorization of the path integral integrand. In this case, complete positivity follows from the fact that 
 \begin{equation}\label{eq: poscond2Appendfield}
 \begin{split}
 & \int \mathcal{D} \phi_{i+1}^+ \mathcal{D} \phi^-_{i+1} \mathcal{D} \phi_i^+ \mathcal{D} \phi^-_i \mathcal{D} z_i  \ \times \\
 & \sum_{\gamma} \int d\vec{x} c^{\gamma}_{i+1,i}(x) \big(v_{i+1,i}^+ (L^{+}_{\gamma})_{i+1,i}(x) e^{\mathcal{I}^+_{i+1,i}} \big) \big(  L^{- }_{\gamma})_{i+1,i}(x)  v_{i+1,i}^{-} e^{\mathcal{I}^-_{i+1,i}} \big)^*     e^{-\mathcal{I}_{C,i+1,i}} \cqstate_i ,
 \end{split}
\end{equation}
is positive when $c^{\gamma} \geq 0$ and $\rho_i$ is a positive density matrix.

\section{Showing the natural class of CQ dynamics is CP}\label{sec: protoactioncp}
In this section, we show that the dynamics defined by Equation \eqref{eq: fieldConfigurationSpace2} takes the form of Equation \eqref{eq: positiveCQ} and is hence completely positive. Since the purely quantum Lagrangian terms appearing in Equation \eqref{eq: positiveCQ} are manifestly CP, we shall focus on the CQ interaction term 
\begin{equation}\label{eq: CQinteractionpos}
  -\frac{1}{2}  \int dx \Delta X^i(q,x) D_{0,ij}(\q,x) \Delta X^j (x) - \frac{1}{2} \int dx   \bar{X}^i(q,x) D_{2, ij}^{-1}(\q,x) \bar{X}^j (q,x),
\end{equation}
where we use the shorthand notation $ X^i(q,x)= \frac{\delta W_{CQ}}{\delta q_i(x)} $, which we assume is Hermitian since it is generated by a real proto-action $W_{CQ}$. For ease of presentation, we will here suppress any potential $q,x$ dependence from $X^i,D_0,D_2$, but these can be added back in.

Expanding Equation \eqref{eq: CQinteractionpos}, we can group terms according to $D_0, D_2^{-1}$ as
\begin{equation}\label{eq: grouptermsApp}
    \begin{split}
     &    -\frac{1}{8} \int dx  d\vec{y } ( 4D_{0,ij} + D_{2,ij}^{-1})((X^+)^i(X^+)^j +  (X^-)^i(X^-)^j) \\
     & +\frac{1}{8} \int d x ( 4D_{0,ij} - D_{2,ij}^{-1})((X^+)^i(X^-)^j+  (X^-)^i(X^+)^j). 
    \end{split}
\end{equation}
We see that the first line in Equation \eqref{eq: grouptermsApp} is of the form $I_{CQ}^+ + (I_{CQ}^-)^* $ and so adheres to the form in Equation \eqref{eq: positiveCQ}. If the trade-off is saturated, this completes the proof that \eqref{eq: fieldConfigurationSpace2} takes the form of Equation \eqref{eq: positiveCQ}.  When it is not saturated, we can write $4D_0-D_2 = c \succeq 0$, where $c_{ij}(q,x)$ is a real, symmetric positive semi-definite matrix. We can then expand the second line of Equation \eqref{eq: grouptermsApp} as 

\begin{equation}\label{eq: remainder}
     \frac{1}{8} \int  d x\  c_{ij}(q,x)((X^+)^i(X^-)^j +  (X^-)^i(X^+)^j), 
\end{equation}
 which, after diagonalizing $c_{ij}$, takes the form of Equation \eqref{eq: positiveCQApp}, and hence defines CP dynamics whenever the condition $4D_0-D_2 = c \succeq 0$ is satisfied.

When the trade-off $4D_0-D_2 = c \succeq 0$ is saturated, i.e., when $c=0$, we can reproduce the full action for the gravity theory of Equation \eqref{eq: fieldConfigurationSpace2}
 \begin{equation}\label{eq:PQG-action}
\begin{split}
    &  \mathcal{I}[ \xql,  \xqr,  g_{\mu \nu}] =    \int dx \bigg[   i\mathcal{L}_{KG}^{\rind}  - i\mathcal{L}_{KG}^{\lind}     -\frac{\det(-g)}{8}( T^{\mu \nu + } - T^{\mu \nu - }) D_{0,\mu \nu \rho \sigma}(T^{\rho \sigma+ } - T^{\rho \sigma- }) \\
    & - \frac{\det(-g)}{128 \pi^2} ( G^{\mu \nu} - \frac{1}{2}( 8 \pi (T^{ \mu \nu })^{\rind} + 8 \pi  (T^{ \mu \nu })^{\lind}  ) D_{0, \mu \nu \rho \sigma}[g] ( G^{\rho \sigma} - \frac{1}{2}( 8 \pi (T^{ \rho \sigma })^{\rind} + 8 \pi  (T^{ \rho \sigma })^{\lind}  )
    \bigg],
    \end{split}
\end{equation}
which we have now verified takes the form of Equation \eqref{eq: positiveCQ}.

\section{Perturbative methods of calculating correlation functions}\label{sec: peturbative}
In this section we study a simple model of CQ interaction to illustrate how one can use standard perturbative methods to calculate classical-quantum correlation functions via CQ Feynman diagrams. 
\new{In the main body, we considered the path integral which constructs a CQ state at a time $t_f$ from a CQ state at time $t_i$. In computing correlation functions of classical-quantum observables, the final state is not important, and so we can perform an integral over all final states of the field $q_f$ at $t_f$ to arrive at the partition function }
\begin{equation}
    Z= \int dq_f \Tr{}{\cqstate(q_f,t_f)}
\end{equation}
which for the configuration space path integral takes the form 
\begin{equation}
  Z=   \int \N \mathcal{D} \xql  \mathcal{D} \xqr   \mathcal{D}\q \ e^{ \mathcal{I}( \xql,  \xqr, \q, t_i,t_f)}\cqstate(q_i,\xqr_i,\xql_i, t_i),
\end{equation}
where now there are no final boundary conditions imposed on the path integral. 

Formally, we can calculate correlation functions by inserting sources $J^+, J^-, J_q$ \new{for the respective fields $\phi^+$, $\phi^-$, $q$} into the path integral, and taking functional derivatives with respect to the sources. The partition function of interest is therefore
\begin{equation}
    Z[J^+,J^-,J_q] = \int\N  \mathcal{D} \xql  \mathcal{D} \xqr   \mathcal{D}\q \ e^{ \mathcal{I}( \xql,  \xqr, \q, t_i,t_f) -i J_+ \phi^+ + iJ^- \phi^- -J_q q}\cqstate(q_i,\xqr_i,\xql_i, t_i).
\end{equation}
In general, the form of the path integral depends on the initial CQ state $\cqstate(q_i,\xqr_i,\xql_i, t_i)$ and any calculation of correlation function must be performed on a case by case basis depending on the initial state. 

However, often we are interested in stationary states, and we would like to obtain information on correlation functions over arbitrary long times by taking the limit $t_i \to -\infty, t_f \to \infty$. In open systems, as well as when calculating scattering amplitudes, it is often assumed that the initial state in the infinite past does not affect the stationary state of the system so that there is a complete loss of memory of the initial state \cite{Sieberer_2016}. Under this assumption, it is possible to ignore the  boundary term containing the initial CQ state $\cqstate(q_i,\xqr_i,\xql_i, t_i)$ and we arrive at the partition function 
\begin{equation}\label{eq: CQpartitionfunction}
   Z[J^+,J^-,J_q] = \int \N \mathcal{D} \xql  \mathcal{D} \xqr   \mathcal{D}\q \ e^{ \mathcal{I}( \xql,  \xqr, \q, -\infty,\infty) -i J_+ \phi^+ + iJ^- \phi^- -J_q q}.
\end{equation}
Using equation \eqref{eq: CQpartitionfunction}, we can then use standard perturbation methods for computing correlation functions in CQ theories. 

As a simple example, \new{consider the CQ theory of Eq. \eqref{eq: Linvariantaction}, but in zero spatial dimensions, so that} the proto-action is given by
\begin{equation}
   W_{CQ}= - \frac{m_q^2 q^2}{2} - \frac{\lambda q^2 \phi^2}{2},
\end{equation}
and the pure quantum action given by 
$S_Q = -\frac{m_{\phi}^2\phi^2}{2}$. Assuming the decoherence diffusion trade-off is saturated, \new{we arrive at the total action via the procedure outlined in the section on the natural class of CQ dynamics (analogous to  Eq. \eqref{eq: decoLor}-\eqref{eq: diffLor}):}
\begin{equation}\label{eq: toyCQexact}
   \mathcal{I}[\phi^{\pm},q] = -\frac{i}{\hbar} \frac{m_{\phi}^2(\phi^{+})^2}{2} + \frac{i}{\hbar}\frac{m_{\phi}^2(\phi^{-})^2}{2} -\frac{1}{2 D_2} \left(   q^2 m_q^4 +  \frac{1}{2}\lambda^2 q^2 ((\phi^{+})^4 + (\phi^{-})^4)  + \frac{1}{2} \lambda q m_q^2 ((\phi^{+})^2 + (\phi^{-})^2)\right).
\end{equation}
\new{The first interaction in the brackets produces a $(q\,\phi^{2})$ three-vertex, while the second yields a $(q^{2}\phi^{4})$ six-vertex; both will show up explicitly in the Feynman rules below.}
We see from Equation \eqref{eq: toyCQexact} that $D_2$ in an interacting CQ theory plays exactly the same role as $\hbar$ in an interacting quantum theory, \new{in the sense that to compute correlation functions, we can work perturbatively in $D_2$.} Note, the double limit $D_2 \to 0, D_2^{-1}\lambda \to 0$ defines a deterministic quantum theory with no classical back-reaction.

We define the free theory as the action independent of any CQ back-reaction
\begin{equation}
   I_{free} = iS^+ - iS^-  -I_{C} =  -i \frac{m^2_{\phi}(\phi^{+})^2 }{2 \hbar} + i \frac{m^2_{\phi}(\phi^{-})^2 }{2 \hbar} -\frac{1}{2 D_2}    q^2 m_q^4.
\end{equation}
 Inserting sources, we find the partition function of the free theory
\begin{equation}
    Z_{free}[J_+,J_-,J_q] = \int d \phi^{\pm} dq e^{I_{free} - \frac{i}{\hbar}J_+ \phi^+ +\frac{i}{\hbar} J_- \phi^- -\frac{1}{D_2} J_q q },
    \label{eq: beforeintegration}
\end{equation}
which can be performed exactly by performing each Gaussian integral individually
\begin{equation}\label{eq: allintergrals}
 Z_{free}[J_+,J_-,J_q] = (\int d \phi^{+} e^{-i\frac{m^2_{\phi}(\phi^{+})^2}{2 \hbar} - iJ_+ \phi^+
} ) (\int d \phi^{-} e^{+i\frac{m^2_{\phi}(\phi^{-})^2}{2 \hbar} + iJ_- \phi^-
} ) (\int d q e^{-\frac{1}{2 D_2}    q^2 m_q^4 - J_q q
} ).
\end{equation}
 Equation \eqref{eq: allintergrals} is evaluated as  
\begin{align}\label{eq: freepartitionexact}
  Z_{free}[J_+,J_-,J_q] & = ( \frac{-2 \pi i \hbar }{m_{\phi}^2}) e^{\frac{i J_+^2}{2 \hbar  m_{\phi}^2}}  ( \frac{2 \pi i \hbar }{m_{\phi}^2}) e^{\frac{-i J_-^2}{2 \hbar  m_{\phi}^2}}  ( \frac{\pi D_2 }{m_{q}^4}) e^{\frac{ J_q^2}{2 D_2 m_q^4 }} \nonumber\\
  &
  :=Z_0 e^{\frac{i J_+^2}{2 \hbar  m_{\phi}^2}} e^{\frac{-i J_-^2}{2 \hbar  m_{\phi}^2}}  e^{\frac{ J_q^2}{2 D_2 m_q^4 }}.
\end{align}
\new{where we have defined $Z_0$ as the interaction free partition function without sources.
From Equation 
\eqref{eq: allintergrals}
we can read out the propagators for the free theory as we would in standard quantum theory --each one can be found by taking two functional derivatives of the corresponding source, while keeping track of the coefficients of $i/\hbar$ and $1/D_2$ from Eq \eqref{eq: beforeintegration}. This gives}
\begin{equation}
    \langle \phi^+ \phi^+ \rangle = - \frac{i  \hbar}{ m_{\phi}^2}, \  \langle \phi^- \phi^- \rangle =  \frac{i  \hbar}{ m_{\phi}^2}, \ \langle q q \rangle = \frac{D_2}{  m_{q}^4} ,
\end{equation}
and we can represent each of the propagators by the following Feynman diagrams
\begin{align}
\label{1PI2pt}
\begin{split}
\diagramone
\end{split}
\end{align}

The full partition function with the CQ interaction turned on then takes the form
\begin{equation}
  Z[J_+,J_-,J_q]  =\langle e^{\mathcal{I}_{CQ}} \rangle = \langle e^{-\frac{1}{2 D_2} \left(   \frac{1}{2}\lambda^2 q^2 ((\phi^{+})^4 + (\phi^{-})^4)  + \frac{1}{2} \lambda q m_q^2 ((\phi^{+})^2 + (\phi^{-})^2)\right)} \rangle 
\end{equation} and we can perform an asymptotic expansion of the CQ interaction in terms of $D_2$ to arrive at the usual Feynman rules for computing correlation functions. Specifically, for terms in the action like $ \lambda_{nml} \phi^{n}_+ \phi^{m}_- q^l$ \new{that is, $n$ copies of
$\phi^+$, $m$ copies of $\phi^-$ and $l$ copies of $q$,
the corresponding Feynman rule assigns to a single
vertex with those $n+m+l$
legs},
	 the  factor
 $\lambda_{nml} n! m! l!$ to each topologically distinct diagram. \new{The extra factorials are the usual symmetry factors: they count all distinct ways of attaching the identical external legs to that vertex, so you do not have to divide by additional combinatorial numbers when summing over diagrams.}

As an example, the CQ interaction term $q (\phi^{\pm})^2$ in Equation \eqref{eq: toyCQexact} has two tri-verticies with strength $-\frac{2!}{4 D_2} \lambda m_q^2 $ and can be represented by the diagrams
\begin{align}
\label{1PI2pt}
\begin{split}
\interactiondiagram
\end{split}
\end{align}
We also have the sextic $q^2 (\phi^{\pm})^4$ interaction with vertex value $-\frac{\lambda^2 4! 2! }{4D_2 }$ which is assigned to each of the following diagrams
\begin{align}
\label{1PI2pt}
\begin{split}
& \interactiondiagramsix \\
& \interactiondiagramsixdash .
\end{split}
\end{align}
\new{Because the purely quantum part of the CQ action is similar to the Schwinger-Keldish path integral, it is often convenient to change basis to the  combinations $\bar{\phi}=\frac{1}{2}\left(\phi^{+}+\phi^{-}\right)$ and $2\Delta\phi=\phi^{+}-\phi^{-}$.  These are typically called the''classical'' field and ''quantum'' field respectively -- terminology that will no doubt be confusing if used in the present context. In this basis, the causal structure of propagators is explicit and only vertices with an odd number of “quantum’’ legs contribute to connected correlation functions, exactly as in the standard Keldysh technique.  Readers who prefer that basis can translate the foregoing formulas straightforwardly.}

\color{black}
\section{Ensuring the CQ path integral is normalized}\label{app: normalizing}
In this section, we show that the CQ action defined by Equation \eqref{eq: fieldConfigurationSpace2} is normalized so long as it contains appropriate classical and quantum kinetic terms. To see the problem of normalization of CQ path integrals in more detail, we will review how the normalization of quantum states occurs in Lindbladian path integrals with a Feynman-Vernon action \cite{FeynmanVernon}, and how probabilities are conserved in higher-derivative classical path integrals. Let us first consider higher-order classical path integrals. We refer the reader to \cite{UCLPILONG} for a complete derivation of normalized CQ path integrals from master equations. 

\subsection{Normalization of higher derivative classical path integrals}
When considering a classical path integral that contains higher derivatives, we should treat $q, \dot{q}$ as independent variables. This is outlined in detail in \cite{Hawking:2001yt}. To that end, we will show how the normalization of the path integral
\begin{equation}\label{eq: classPI}
    \p(q_f, \dot{q}_f,t_f) = \int^{B} \ \mathcal{D}q e^{-\int_{t_i}^{t_f} dt[\ddot{q}- f(\dot{q}, q)  ]^2}p(q_i, t_i)
\end{equation}
occurs. In Equation \eqref{eq: classPI}, note that the boundary conditions are given by $B=\{q(t_f) = q_f, \dot{q}(t_f) = \dot{q}_f\}$, which involve both  $q$ and  $\dot{q}$.

To check normalization, we consider Equation \eqref{eq: classPI} for small $\delta t$, with $t_n = \delta t + t_{n-1}$
\begin{equation}\label{eq: shorttimeclass}
\begin{split}
   &  \p(q_{n+1}, q_{n+2},t_{n+1}) =  \int dq_{n} e^{-\delta t[\frac{q_{n+2}-2q_{n} + q_{n+1}}{\delta t}-f(q_{n+1}, q_n) ]^2}   p(q_n, q_{n+1}, t_n).
    \end{split}
\end{equation}
The norm of the probability distribution is found by performing the integral over the final variables $q_{n+1}, q_{n+2}$
\begin{equation}\label{eq: highernorm}
  \begin{split}
 & \int dq_{n} dq_{n+1} dq_{n+1}  e^{-\delta t[\frac{q_{n+2}-2q_{n} + q_{n+1}}{\delta t}-f(q_{n+1}, q_n) ]^2}  \times p(q_n, q_{n+1}, t_n).
\end{split}  
\end{equation}
Equation \eqref{eq: highernorm} defines a standard Gaussian integral over the $q_{n+2}$ coordinate. Hence, the $q_{n+2}$ integral eats the action up to a Gaussian normalization factor that we can calculate exactly, and we are left with 
\begin{equation}
1=  \int dq_{n} dq_{n+1} \N p(q_n, q_{n+1}, t_n),
\end{equation}
so we can simply absorb $\N$ into the measure, and the path integral will be normalized. If we were to include a $q$ dependent diffusion coefficient $D_2(q,\dot{q})$ in Equation \eqref{eq: classPI}, then the Gaussian integral will be $q$ dependent, and this will need to be included in the measure for $\mathcal{D}q$ \cite{UCLPILONG}. The important point is that the higher derivative terms in the classical path integral are standard Gaussian integrals if we consider $q$ and $\dot{q}$ as independent variables. Hence, by including kinetic terms in the classical part of the action we expect that the classical contribution to the path integral defined by Equation \eqref{eq: fieldConfigurationSpace2} can be normalized to give conserved probabilities. We show this explicitly in Section \ref{sec: CQnormalizationApp}.

\subsection{Normalization of Feynman-Vernon path integrals}
Let us now consider a Feynman-Vernon quantum path integral with a decoherence term. Consider first the path integral for a quantum state $\sigma$
\begin{equation}\label{eq: lind0}
\begin{split}
     & \sigma(\phi^+_f, \phi^-_f, t_f) = \int^B \mathcal{D} \phi^+ \mathcal{D}\phi^-  e^{\int_{t_i}^{t_f}dt i[\dot{\phi_+}^{2} + V(\phi_+)] -i[\dot{\phi_-}^{2} + V(\phi_-)]  -\frac{D_0}{2}(L(\phi_+)-L(\phi_-))^2} \sigma(\phi_i^+, \phi^-_i, t_i),
    \end{split}
\end{equation}
where $B$ imposes the final state boundary conditions on the bra and ket fields, and $L(\phi)$ is an arbitrary operator of $\phi$ but not of its derivatives. 

For Equation \eqref{eq: lind0}, it will prove insightful to show how the kinetic term enforces the normalization of the quantum state. To that end, consider the short time version of Equation \eqref{eq: lind}
\begin{equation}\label{eq: lind}
\begin{split}
      \sigma(\phi^+_{n+1}, \phi^-_{n+1}, t_f)  &= \int d\phi^+_n d\phi^-_n    e^{\delta t [i(\frac{\phi^+_{n+1} -\phi^+_n }{\delta t})^{2} + iV(\phi^+_n) -i(\frac{\phi^-_{n+1} -\phi^-_n }{\delta t})^{2} -i V(\phi^-_n)] }\\
    & \times e^{ -\delta t \frac{D_0}{2}(L(\phi_n^+)-L(\phi_n^-))^2}\sigma(\phi_n^+, \phi^-_n, t_n).
    \end{split}
\end{equation}
The trace of the quantum state is found by matching the $\phi^+_{n+1}= \phi^-_{n+1} = \phi$ fields and integrating over $\phi$
\begin{equation}
\begin{split}
      \int d\phi^+_n d\phi^-_n d\phi  e^{\frac{i}{\delta t} \phi( \phi_n^+- \phi_n^-)} e^{i \delta t [V(\phi_n^+) -i V(\phi_n^-)] } e^{ -\delta t \frac{D_0}{2}(L(\phi_n^+)-L(\phi_n^-))^2}\sigma(\phi_n^+, \phi^-_n, t_n).
    \end{split}
\end{equation}
Performing the integration over $\phi$ gives rise to a delta function $\delta(\phi^+_n -\phi^-_n)$. Hence, the quantum state is normalized to constant factors that can be absorbed.

However, had we included higher-order kinetic terms in the decoherence sector, we would not have found this normalization. In particular, if the decoherence term was instead
\begin{equation}
\begin{split}
   \int_{t_i}^{t_f}dt \frac{1}{2}D_0( \dot{\phi}_+^2 - \dot{\phi}_-^2)^2,
    \end{split}
\end{equation}
then the delta function integral is not imposed, and the state is no longer normalized to constant factors.

As such, for the path integral to be normalized with higher derivative decoherence terms, one needs to also add higher derivative kinetic terms in the action. In this case, the action 
\begin{equation}
\begin{split}
    S &= \int_{t_i}^{t_f}dt i[\dot{\phi_+}^{2} + \ddot{\phi_+}^{2} - V(\phi_-)] -i[\dot{\phi_-}^{2} + \ddot{\phi_-}^{2} - V(\phi_-)] - \frac{1}{2}D_0( \dot{\phi}_+^2 - \dot{\phi}_-^2)^2
    \end{split}
\end{equation}
is normalized up to constant factors by the same argument, so long as we treat $\phi $ and $\dot{\phi}$ as independent variables to be specified in the quantum state; this is also argued for independent reasons in \cite{Hawking:2001yt}. To see this, one does the short time expansion, treating $\phi $ and $\dot{\phi}$ as independent variables as in the higher derivative classical path integral. Computing the trace then sets $\phi_{n+2}^{\pm}$ equal to each other, as well as the setting the $\phi_{n+1}^{\pm}$ fields equal to be the same. The $\phi_{n+2}$ integral then enforces a delta function over $\delta(\phi^+_n-\phi^-_n)$, which kills the decoherence term and means that the path integral is normalized up to constant factors. 
\subsection{Normalization of CQ path integrals}\label{sec: CQnormalizationApp}
In this section, we show that any CQ path integral with action 
\begin{equation}\label{eq: normalizedCQmainLowerAppendix}
\begin{split}
     I[q,\phi^+, \phi^-]&= \int dt i \dot{\phi}_+^2 + i V(\phi^+) - i \dot{\phi}_-^2 -i V(\phi^-) - \frac{D_0(q,\dot{q}, \phi^+)}{2}(\ddot{q} + f(q,\dot{q}, \phi^+))^2 \\
    & - \frac{D_0(q,\dot{q}, \phi^-)}{2}(\ddot{q} + f(q,\dot{q}, \phi^-))^2,
    \end{split}
\end{equation}
is normalized up-to constant factors when $D_0>0$. In the case where $D_0$ has a functional dependence on the fields one must make sure to also include a factor of  $\sqrt{\det(D_0(q,\dot{q}, \phi)}$ in the path integral measure. 
We further show that any higher-derivative action 
\begin{equation}\label{eq: normalizedCQApp}
\begin{split}
     I[q,\phi^+, \phi^-]&= \int dt i \ddot{\phi}_+^2 + i V(\phi^+, \dot{\phi}^+) - i \ddot{\phi}_-^2 -i V(\phi^-, \dot{\phi}^-) \\
    &- \frac{D_0(q,\dot{q}, \phi^+, \dot{\phi^+})}{2}(\ddot{q} + f(q,\dot{q}, \phi^+, \dot{\phi}^+))^2  - \frac{D_0(q,\dot{q}, \phi^-, \dot{\phi^-})}{2}(\ddot{q} + f(q,\dot{q}, \phi^-, \dot{\phi}^-))^2
    \end{split}
\end{equation}
is also normalized. Equation's \eqref{eq: normalizedCQmainLowerAppendix} and \eqref{eq: normalizedCQApp}, are generic type of action one gets from varying Equation \eqref{eq: fieldConfigurationSpace2} with a CQ proto action that has second order equations of motion for the classical degree of freedom. 

The steps in showing Equation \eqref{eq: normalizedCQApp} follow in the same way as the discussions of classical and quantum path integrals. Firstly, because the action is higher derivative, the CQ state is specified through $\cqstate(q,\dot{q}, \phi^{\pm}, \dot{\phi}^{\pm})$. 

Taking the trace at the $t_{n+1} = t_{n}+ \delta t$ time-step therefore involves identifying $\phi_{n+2}^+=\phi_{n+2}^- = \phi_{n+2}$ and $\phi_{n+1}^+=\phi_{n+1}^-=\phi_{n+1}$. We then integrate over the $\phi_{n+2}$ and $\phi_{n+1}$ variables, as well as over the $q_{n+2}, q_{n+1}$ classical degrees of freedom.

Let us first look at the higher derivative quantum kinetic term. This can be expanded as
\begin{equation}
\begin{split}
    \ddot{\phi}_+^2 -\ddot{\phi}_-^2 & \sim   (\phi_{n+2} - 2\phi_n^+ + \phi_{n+1})^2 - (\phi_{n+2} - 2\phi_n^- + \phi_{n+1})^2\\
   & = 4\big[ (\phi_{n}^+)^2-(\phi_{n}^-)^2 + \phi_{n+2}( \phi_n^--\phi_n^+) + \phi_{n+1}(\phi_n^- -\phi_n^+) \big].
    \end{split}
\end{equation}
Hence, integrating over $\phi_{n+2}$ gives a delta function in $\delta(\phi_n^--\phi_n^+)$. As a consequence of this, all the bra and ket fields in the path integral are identified. We are therefore left with the action 
\begin{equation}\label{eq: gaussian_final_cq}
\begin{split}
    & I'[q,\phi ]= -\int dt D_0(q,\dot{q}, \phi,\dot{\phi})(\ddot{q} + f(q,\dot{q}, \phi, \dot{\phi}))^2 .
    \end{split}
\end{equation}
Since all the bra and ket quantum fields are identified, normalization of Equation \eqref{eq: normalizedCQApp} is equivalent to ensuring that Equation \eqref{eq: gaussian_final_cq} is normalized.

As we saw for the classical path integrals, integrating Equation \eqref{eq: gaussian_final_cq} over the $\ddot{q}$ at second time step implements a standard Gaussian integral. If $D_0$ is dependent on the fields, we therefore pick up a term $(\sqrt{\det(D_0(q,\dot{q}, \phi)})^{-1/2},
$
which we must cancel in the measure by including a $\sqrt{\det(D_0(q,\dot{q}, \phi)}$ term, as in \cite{UCLPILONG}. It can also be exponentiated into the action by introducing Bosonic and Fermionic Faddeev-Poppov fields \cite{Bastianelli_2017}. This determinant term commonly arises in the study of Fokker-Plank type equations when the noise is multiplicative \cite{Onsager1953Fluctuations,Dekker, Bastianelli_2017}. With this in mind, once we have integrated over $\ddot{q}$, the action vanishes and we are left with the normalization of the initial CQ state. Hence the path integral preserves the normalization of CQ states.

In a similar manner, we can also show that the path integral of Equation \eqref{eq: normalizedCQmainLowerAppendix}. To see this, we first take the trace of the system, setting $\phi_{n+1}^+= \phi_{n+1}^- =\phi$. Integrating over $\phi$ then enforces a delta function $\delta( \phi^+-\phi^-)$. We are then left with the action 
\begin{equation}\label{eq: gaussian_final_cq_lower}
\begin{split}
    & I'[q,\phi ]= -\int dt D_0(q,\dot{q}, \phi)(\ddot{q} + f(q,\dot{q}, \phi))^2, 
    \end{split}
\end{equation}
and we can again perform the Gaussian integral over $\ddot{q}$ to arrive at a normalized path integral if $\sqrt{\det(D_0(q,\dot{q}, \phi)}$ is included in the measure.

\end{document}